\documentstyle[aps]{revtex}

\begin{document}
\draft
\title{NONLINEAR DISCRETE SYSTEMS WITH NONANALYTIC\\ DISPERSION
RELATIONS\thanks{Work supported in part by INFN and by MURST}}
\author{M. BOITI$^{\dag}$, J. LEON, F. PEMPINELLI\thanks{Permanent
address: Dipartimento di Fisica, Universit\`{a} di Lecce, I-73100 LECCE} \\
Physique Math\'ematique et Th\'eorique, CNRS-URA 768\\
34095 MONTPELLIER Cedex 05 (FRANCE)}
\date{ }\maketitle

\begin{abstract}
A discrete system of coupled waves (with nonanalytic dispersion relation) is
derived in the context of the spectral transform theory for the Ablowitz
Ladik spectral problem (discrete version of the Zakharov-Shabat system).
This 3-wave evolution problem is a discrete version of the {\em stimulated Raman
scattering} equations, and it is
shown to be solvable for arbitrary boundary
value of the two {\em radiation fields} and initial value of the {\em medium
state}. The spectral transform is constructed on the basis of the $\overline{%
\partial}$-approach.
\end{abstract}

\section{Introduction}

This paper relates the study of the following discrete coupled system for
the three fields $A_1(\theta ,n,t)$, $A_2(\theta ,n,t)$ and $q(n,t)$
\begin{eqnarray}
&&A_1(\theta ,n,t)-A_1(\theta ,n{-}1,t)=e^{-in\theta }q(n,t)A_2(\theta ,n,t)
\nonumber \\
&&A_2(\theta ,n,t)-A_2(\theta ,n{-}1,t)=-e^{in\theta }\overline{q}%
(n,t)A_1(\theta ,n,t)  \label{basic-sys} \\
&&q_t(n,t)=\frac{\rho (n,t)}{2\pi }\int_{-\pi }^{+\pi }d\theta \,e^{in\theta
}(A_1*A_2)(\theta,n,t)  \nonumber
\end{eqnarray}
where $\theta \in [-\pi ,+\pi ]$, $n\in {\bf Z}$ and $t\in {\bf R}^{+}$. The
interaction term here above is defined as the {\em coupling} factor
\begin{eqnarray}
&&(A_1*A_2)(\theta ,n,t)=  \nonumber \\
&&\quad g(\theta ,t)A_1(\theta ,n-1,t)\overline{A}_2(\theta ,n,t)+\overline{%
g}(\theta ,t)A_1(\theta ,n,t)\overline{A}_2(\theta ,n-1,t)  \label{bra-ket}
\end{eqnarray}
where $g(\theta ,t)$ is an arbitrary function in $L^2([-\pi ,+\pi ])$ (which
could also be time dependent), and where the {\em energy ratio} $\rho (n,t)$
at the site $n$ (it will be shown that this quantity is actually $\theta $%
-independent) is defined as
\begin{equation}
\rho (n,t)={\frac{|I_1(\theta ,t)|^2+|I_2(\theta ,t)|^2}{|A_1(\theta
,n,t)|^2+|A_2(\theta ,n,t)|^2}}  \label{rho-def}
\end{equation}
for the following definition of the boundary values $I_1(\theta ,t)$ and $%
I_2(\theta ,t)$ (input data)
\begin{equation}
I_1=\displaystyle{\mathop{\mbox{lim}}_{n\to +\infty }}A_1(\theta ,n,t),\quad
I_2=\displaystyle{\mathop{\mbox{lim}}_{n\to +\infty }}A_2(\theta ,n,t).
\label{bound-val-def}
\end{equation}

One of the main results is the proof that the system (\ref{basic-sys}) with
data ({\em initial-boundary value problem})
\begin{equation}
q(n,0),\quad I_1(\theta,t),\quad I_2(\theta,t)  \label{data-bound}
\end{equation}
is integrable. In particular this work provides the first instance of an
integrable nonlinear discrete system with {\em nonanalytic dispersion
relation}.

An interesting limit of the above equation arises when the arbitrary
distribution $g(\theta )$ goes to a Dirac delta function, for instance $%
\delta (\theta )$. Then the system reads (now $A_j(n,t)$ denotes $A_j(\theta
,n,t)|_{\theta =0}$)
\begin{eqnarray}
&&A_1(n,t)-A_1(n{-}1,t)=q(n,t)A_2(n,t)  \nonumber \\
&&A_2(n,t)-A_2(n{-}1,t)=-\overline{q}(n,t)A_1(n,t)  \label{basic-sys-sharp}
\\
&&q_t(n,t)=\rho (n,t)[A_1(n{-}1,t)\overline{A}_2(n,t)+A_1(n,t)\overline{A}%
_2(n{-}1,t)]  \nonumber
\end{eqnarray}
and it is called the {\em sharp line limit} of (\ref{basic-sys}). Although a
definite physical application of such an equation does not exist by now, it
still can be understood in the following way. $A_1$ and $A_2$ are the two
envelopes of some high frequency (HF) oscillations (say at frequency $\omega
_1$ and $\omega _2$) which interact resonantly on each site $n$ with a
medium constituted of oscillators of envelope $q(n)$ and frequency $\Omega
=\omega _1-\omega _2$, with a coupling intensity proportional to the
relative amount $\rho (n)$ of the {\em total HF energy} which has reached
the site $n$. Then the physical data are the input values $A_{1,2}(\theta
,n,t)$ at $n=\infty $ of the HF external excitations, and the initial state
of the medium oscillators.

The method used to build and solve (\ref{basic-sys}) is the {\em inverse
spectral\/} (or {\em scattering}) {\em transform} (IST) well known also as
the {\em nonlinear Fourier transform} \cite{bookist}. Indeed the method, in
its principle, works like a Fourier transform. It associates to the field,
solution of a nonlinear evolution equation, its {\em spectral transform}
which evolves linearly. Then the field at time $t$ is reconstructed from the
spectral transform at time $t$ by solving the {\em inverse spectral
problem}. This method has been widely studied and extended to various
interesting nonlinear evolution problems. We are particularly interested in
three types of extension which will be used all three together.

The first extension involved here is the use of {\em discrete} spectral
problems to solve discrete (in space) nonlinear evolution equations. Famous
instances of integrable discrete systems are the Toda lattice \cite{toda}
and a special discrete version of the nonlinear Schr\"{o}dinger equation
which has been proposed and integrated by Ablowitz and Ladik by using a
discrete version of the Zakharov--Shabat spectral problem \cite{abladi}.
This so called Ablowitz--Ladik spectral problem and the related nonlinear
differential--difference equations with polynomial dispersion relations have
been extensively studied (see \cite{Kako77} \nocite{Kako79} \nocite{Chiu}
\nocite{Levi78}\nocite{Bruschi79} \nocite{Levi80} \nocite{Khabibulin} \nocite
{Novokshenov} \nocite{Kulish} \nocite{gerdj} \nocite{Herbst}\nocite{Zhang}
\nocite{vek-kon}-- \cite{Konotop}). Recently a different version of the
discrete Zakharov-Shabat system has been proposed in order to keep the
canonical Poisson structure of the continuous case \cite{merol-Ragnis}.

The second domain considered here concerns the extension of the spectral
transform to the case of {\em nonanalytic dispersion relations}
\cite{kaunew}\cite{jl-1}. The first instance of such an integrable system is the
self-induced transparency (SIT) equations of McCall-Hahn \cite{machah} which
was shown to possess a Lax pair in \cite{lamb},
was given a N-soliton solution in \cite{gibbon}
and later studied and completely solved in \cite{abkane}.
These systems generically describe wave-wave interactions for
which some boundary value are prescribed. These boundary values are strongly
dependent on the physical problem under consideration. For instance the
problem of superfluorescence in two-level media results in the same equation
as SIT but with different boundary values and consequently quite different
generic properties resulting mainly from a linear but {\em non homogeneous}
evolution of the spectral transform \cite{gabzak}.

The third extension used is the generalization of the solution of an
evolution equation with a nonanalytic dispersion relation to the case of
{\em arbitrary boundary values} \cite{leonsing}. In this case, the evolution
of the spectral transform can be not only non homogeneous but also {\em %
nonlinear} and still has interesting physical application. In particular the
problem of stimulated Raman scattering of a high energy long Laser pulse in
a gas has been solved by this technique \cite{jlprl}\cite{jl-fini}.

The paper is organized as follows. In sect. 2 we summarize the method of
solution of the system (\ref{basic-sys}) and provide there only the
resulting formulae.

In sect. 3 the principal Lax operator or, more precisely, the associated
spectral problem (a special reduction of the Ablowitz--Ladik spectral
problem) is used to define the spectral data (or nonlinear Fourier
transform). This is done by selecting the basic set of Jost solutions and
then proving that they obey a Riemann-Hilbert problem in the spectral
parameter.

The inverse spectral problem is then solved in sect. 4 by means of the $%
\overline{ \partial}$-formulation of the spectral problem, which means that
the {\em potentials} are reconstructed from the {\em spectral data}. There
the compatible {\em reductions} are also considered, which will allow in
particular to obtain simpler integrable equations with an easier
interpretation.

The sect. 5 is devoted to the formulation of the inverse spectral transform
on the basis of the $\overline{\partial}$-problem. More precisely, having
previously shown that the spectral problem (for the principal Lax operator)
leads naturally to a $\overline{\partial}$-problem, we prove here the
reverse statement. This is useful in the following for considering the $%
\overline{\partial}$-problem itself as the starting tool.

The general discrete integrable systems with nonanalytic dispersion
relations is then constructed by requiring a time evolution of the spectral
transform with a {\em non-analytic} dispersion law and a {\em non-homogeneous%
} term.

These results are used in sect. 6 to prove that indeed the system (\ref
{basic-sys}) with the arbitrary boundary values (\ref{bound-val-def}) is
solvable. That means that we obtain the time evolution of the spectral data
in terms of the boundary values and the spectral transform of the initial
datum $q(n,0)$. An interesting case corresponds to the growth of the field $%
q(n,t)$ on an initial medium at rest, that is for $q(n,0)\equiv 0$. The
method furnishes in such a case the explicit output values of the HF fields $%
A_1(\theta ,n,t)$ and $A_2(\theta ,n,t)$ for $n=-\infty $.


\section{Solution of the System. A Summary}

The general method to generate solutions of (\ref{basic-sys}) is sketched
hereafter. The starting point is the {\em spectral transform} of the initial
datum $q(n,0)$, namely the set of two scalar functions $\alpha $ and $\beta $
defined on the unit circle, a sequence of N discrete points $k_j$ outside
the unit disc to each of which are associated N complex constants $C_j$.
This set is given at $t=0$ as
\begin{equation}
\alpha (\zeta ,0),\quad \beta (\zeta ,0),\quad \zeta =e^{i\theta };\quad
C_j(0),\quad k_j,\quad |k_j|>1,\quad j=1,..,N.  \label{min-spec-dat}
\end{equation}
In the language of the scattering theory, $\alpha $ is called the {\em %
reflection coefficient}, $\beta $ the {\em transmission coefficient}, $N$
the number of {\em bound states} $k_j$ and $C_j$ the related {\em %
normalization coefficients}. The effective construction of these data from $%
q(n,0)$ is displayed in sect. 3, but here we just consider that the set (\ref
{min-spec-dat}) is given and we show how to build from it a solution of (\ref
{basic-sys}). It is worth mentioning that in the linear limit case of {\em %
small} $q(x,0)$, $\alpha (\zeta ,0)$ becomes the Fourier transform of $q$
(with parameter $2\zeta $), $\beta (\zeta ,0)$ become 1 and all the $C_j$'s
vanish (no discrete spectrum, or else no solitons in the linear limit).

The first step is to construct the {\em spectral transform} at time $t$ by
solving
\begin{eqnarray}
\partial _t\alpha &=&\alpha \frac{g+\overline{g}}2\left( |I_1(\theta
,t)|^2-|I_2(\theta ,t)|^2\right) -(g+\overline{g})\overline{I_1}I_2
\label{S-evol-alpha-ph} \\
&&-\alpha \frac 1{2\pi i}P\oint_{{\cal C}}\frac{\,d\zeta ^{\prime }}{\zeta
^{\prime }-\zeta }(g+\overline{g})(|I_1|^2-|I_2|^2) \\
&&+\alpha \frac 1{2\pi i}\oint_{{\cal C}}\frac{d\zeta ^{\prime }}{\zeta
^{\prime }}\left( g|I_1|^2-\overline{g}|I_2|^2\right) , \\
\partial _tk_j &=&0, \\
\partial _tC_j(t) &=&-C_j(t)\frac 1{2\pi i}P\oint_{{\cal C}}\frac{\,d\zeta
^{\prime }}{\zeta ^{\prime }-k_j}(g+\overline{g})(|I_1|^2-|I_2|^2)
\label{S-disc-spec-evol} \\
&&+C_j(t)\frac 1{2\pi i}\oint_{{\cal C}}\frac{d\zeta ^{\prime }}{\zeta
^{\prime }}\left( g|I_1|^2-\overline{g}|I_2|^2\right) .
\end{eqnarray}
Unlike in the continuum case,  the transmission coefficient $\beta(\zeta ,t)$
cannot be computed directly from $\alpha(\zeta ,t)$ and
it becomes necessary to solve the equation
\begin{equation}\label{S-evol-beta}
\beta (\zeta,t)^{-1}\partial_t\beta(\zeta,t) =-{\frac 12}\,
\frac{\partial_t|\alpha |^2}{1+|\alpha |^2}
+{\frac 1{2i\pi }}P\!\!\oint_{\cal C}
\frac{\zeta}{\zeta^{^{\prime }}}\
\frac{d\zeta ^{^{\prime }}}{\zeta ^{^{\prime }}-\zeta }
\frac{\partial_t|\alpha |^2}{1+|\alpha |^2}.
\end{equation}
In the integrals here above, $P$ denotes the Cauchy principal value and $%
{\cal C}$ the unit circle in the complex plane.

Although not elementary, in the case when $I_1$ and $I_2$ are given
independently of $\alpha $ and $\beta $ the above system of equations is
{\em linear} and can in principle be explicitly solved as soon as the {\em %
initial data} (\ref{min-spec-dat}) and the boundary values (\ref
{bound-val-def}) are known. Before going further, it is already worth
remarking that if the quantity $I_1I_2$ does not vanish, then the evolution
for the reflection coefficient $\alpha (\zeta ,t)$ has a non-homogeneous
term. Consequently the solution can grow on the {\em initial vacuum} $%
q(n,0)=0$ which has the spectral transform
\begin{equation}
\alpha (\zeta ,0)=0,\quad \beta (\zeta ,0)=1;\quad N=0.  \label{vac-spec-dat}
\end{equation}

The second step consists in solving the following system of linear integral
equation for the unknowns $\phi _i(k,n,t)$ for $|k|<1$
\begin{eqnarray}
\left( \matrix{\phi_1(k)\cr \phi_2(k)}\right) &=&\left( \matrix{ 1\cr 0}
\right) +{\frac 1{2i\pi }}\oint_{{\cal C}}{\frac{d\zeta ^{\prime }}{\zeta
^{\prime }-k}}\,{\frac k{\zeta ^{\prime }}}\,(\zeta ^{\prime })^{-n}\,\alpha
(\zeta ^{\prime })\left( \matrix{-\overline{\phi}_2(\zeta')\cr
\overline{\phi}_1(\zeta')} \right) +  \nonumber \\
{\,} &{\,}&+\sum_{j=1}^N{\frac{C_j}{k_j-k}}\,{\frac k{k_j}}(k_j)^{-n}\left( %
\matrix{-\overline{\phi}_2({1/\overline{k}_j})\cr
\overline{\phi}_1({1/\overline{k}_j})}\right) .  \label{sol-a-1-2}
\end{eqnarray}

The solution of (\ref{basic-sys}) then reads (last step) for $\zeta
=e^{i\theta }$
\begin{eqnarray}
&&\left( \matrix{A_1\cr A_2}\right) =I_1\left( \matrix{\phi_1(\zeta)\cr
\zeta^n\phi_2(\zeta)}\right) +I_2\left( \matrix{-\zeta^{-n}\overline\phi_2(
\zeta)\cr \overline\phi_1(\zeta)}\right)  \label{sol-basic} \\
&&q(n{+}1,t) =-\overline{\phi }_2^{(-1)}(n,t)  \label{q-sol-basic}
\end{eqnarray}
with $\phi _2^{(-1)}$ the coefficient of $k^{-1}$ in the Laurent expansion
of the solution $\phi _2(k,n,t)$. This achieves the solution of the
nonlinear system (\ref{basic-sys}) with the arbitrary boundary values (\ref
{bound-val-def}) as a sequence of {\em linear} operations.

An interesting information here is the output values (vs. the input values)
of the fields $A_j$ (values for $n\to -\infty $) which will be proved to be
\begin{equation}
\left( \matrix{A_1\cr A_2}\right) \displaystyle{\mathop{\longrightarrow}
_{n\to -\infty }}\,{\frac 1{1+|\alpha |^2}}\,\left( \matrix{I_1\beta+I_2%
\overline{\alpha}\beta\cr -I_1\alpha \overline{\beta}+I_2\overline{\beta}}%
\right)  \label{output}
\end{equation}
for the input
\begin{equation}
\left( \matrix{A_1\cr A_2}\right) \displaystyle{\mathop{\longrightarrow}
_{n\to +\infty }}\,\left( \matrix{I_1\cr I_2}\right)  \label{input}
\end{equation}
(note the necessity to compute not only the reflection coefficient $\alpha $
at time $t$ but also the transmission coefficient $\beta $). This result,
besides having a physical interest, has the nice property of being {\em %
explicit}. Indeed it does not require solving the integral equations (\ref
{sol-a-1-2}). Actually, when the system (\ref{basic-sys}) is viewed as
describing the interaction of radiation components $A_j$ with matter, the
relevant (measured) physical information is the output values of the
radiation components.

Now we can compute the ratio $\rho (-\infty ,t)$ of transmitted {\em photon
number}, defined in (\ref{rho-def}), as
\begin{equation}
\rho (n,t)\displaystyle{\mathop{\longrightarrow}_{n\to -\infty }}\,{\frac{
1+|\alpha |^2}{|\beta |^2}},  \label{trans-photon}
\end{equation}
while we have obviously
\begin{equation}
\rho (n,t)\displaystyle{\mathop{\longrightarrow}_{n\to +\infty }}\,1.
\label{input-photon}
\end{equation}
This is unlike in the continuous case for which we would find $\rho (-\infty
,t)=1$, and results effectively from the discrete nature of (\ref{basic-sys}%
). Indeed a direct calculation leads to the following {\em total photon
number} non-conservation
\begin{equation}
|A_1(\zeta ,n-1)|^2+|A_2(\zeta ,n-1)|^2=(1+|q(n)|^2)(|A_1(\zeta
,n)|^2+|A_2(\zeta ,n)|^2).  \label{S-non-cons}
\end{equation}
As a consequence we obtain from (\ref{rho-def}) and the above relation
\begin{equation}
\rho ^{-1}(n,t)=\prod_{i=n{+}1}^\infty (1+|q(i,t)|^2)  \label{S-prod-ener}
\end{equation}
which proves in particular that the energy ratio $\rho (n,t)$ indeed does
not depend on the variable $\theta $.


\section{The Spectral Problem}

In the case of the discrete variable, a spectral problem is understood as a
difference equation for some unknown $\mu(n)$, which involves explicitly an
external parameter, the {\em spectral parameter}, and a set of given $n$%
-dependent coefficients, the {\em potentials}. Solving a spectral problem
results in defining the set of {\em spectral data} (functions of the
parameter $k$) in such a way that they are in bijection with the set of
potentials (in some given class of functions). We shall work here in the
space of $2\times2$ matrices and adopt an equivalent form of the reduced
Ablowitz-Ladik spectral problem used in \cite{abladi} to integrate the
discrete nonlinear Schr\"odinger equation. In our case we are able to write
the spectral transform as a $\overline{\partial}$-problem for the matrix $%
\mu(k,n)$, which results in a simple formulation of the inverse problem
together with a very convenient tool for building and solving nonlinear
evolutions, in particular those with nonanalytic dispersion relations and
boundary value data.

\subsection{Equation and Jost solutions}

Let us consider the discrete spectral problem
\begin{equation}
\mu (k,n{+}1)-\Lambda ^{-1}\,\mu (k,n)\Lambda =Q(n{+}1)\mu (k,n{+}1),
\label{spec-pb}
\end{equation}
with the following definitions
\begin{equation}
\Lambda (k)=\left( \matrix{1/z & 0 \cr 0 & z }\right) ,\quad z^2=k,\quad
Q(n)=\left( \matrix{0 & q(n) \cr r(n) & 0 }\right)  \label{Lambda}
\end{equation}
where $k$ is the spectral parameter which belongs to the domain ${\cal D}=
{\bf C}-\{0,\infty \}$. Up to a change of $n\to -n$, $k\to 1/k$ and a
rescaling by convenient powers of $z$ of the matrix elements of $\mu $ it is
just the special reduction of the Ablowitz--Ladik spectral problem
associated to the integrable discrete nonlinear Schr\"{o}dinger equation.
The solution of (\ref{spec-pb}) possesses the property
\begin{equation}
\det \{\mu (k,n-1)\}=\det \{\mu (k,n)\}[1-r(n)q(n)].  \label{det-mu-n}
\end{equation}

The solution of this spectral problem goes through the construction of some
well chosen solutions (the Jost solutions) out of some particular asymptotic
behaviors. These solutions are denoted by $\mu ^{\pm }$ and are defined by
the following discrete integral equations
\begin{eqnarray}
&&\left( \matrix{\mu_{11}^-(k,n) \cr \mu_{21}^-(k,n) }\right) =\left( %
\matrix{1\cr0}\right) +\left( \matrix{-\displaystyle\sum_{i=n+{1}}^{+
\infty}q(i) \mu_{21}^-(k,i)\cr
\displaystyle\sum_{i=-\infty}^{n}k^{i-n}r(i)\mu_{11}^-(k,i)}\right)
\label{mu1-moins} \\
&&\left(\matrix{\mu_{11}^+(k,n) \cr \mu_{21}^+(k,n) }\right)= \left( %
\matrix{1\cr0}\right)+ \left(\matrix{-\displaystyle\sum_{i=n+{1}}^{+
\infty}q(i) \mu_{21}^+(k,i)\cr
-\displaystyle\sum_{i=n+{1}}^{+\infty}k^{i-n}r(i)\mu_{11}^+(k,i)}\right)
\label{mu1-plus} \\
&&\left(\matrix{\mu_{12}^-(k,n) \cr \mu_{22}^-(k,n) }\right)= \left( %
\matrix{0\cr1}\right)+ \left(\matrix{-\displaystyle\sum_{i=n+{1}}^{+
\infty}k^{n-i}q(i)\mu_{22}^-(k,i)\cr
-\displaystyle\sum_{i=n+{1}}^{+\infty}r(i)\mu_{12}^-(k,i)}\right)
\label{mu2-moins} \\
&&\left(\matrix{\mu_{12}^+(k,n) \cr \mu_{22}^+(k,n) }\right)= \left( %
\matrix{0\cr1}\right)+ \left(\matrix{\displaystyle\sum_{i=-\infty}^{n}
k^{n-i}q(i) \mu_{22}^+(k,i)\cr
-\displaystyle\sum_{i=n+{1}}^{+\infty}r(i)\mu_{12}^+(k,i)}\right).
\label{mu2-plus}
\end{eqnarray}
We will make use also of the notation
\begin{equation}
\mu_1^\pm=\left(\matrix{\mu_{11}^\pm \cr \mu_{21}^\pm }\right), \quad
\mu_2^\pm=\left(\matrix{\mu_{12}^\pm \cr \mu_{22}^\pm }\right).
\label{def-vec-mat}
\end{equation}

The above integral equations allow then to obtain, for some given class of
potentials, the analytical properties of the solutions in the domain ${\cal %
D }$ of the complex $k$-plane, for all $n$. The function $\mu_1^+(k,n)$ is
holomorphic inside the unit circle, the function $\mu_2^-(k,n)$ is
holomorphic outside the unit circle, the function $\mu_1^-(k,n)$ is
meromorphic outside the unit circle where it has a finite number $N^-$ of
simple poles $k_j^-$, the function $\mu_2^+(k,n)$ is meromorphic inside the
unit circle where it has a finite number $N^+$ of simple poles $k_j^+$.
Moreover the two solutions $\mu^\pm$ are continuous on the unit circle.

\subsection{Riemann-Hilbert problem and spectral data}

The method to obtain from the integral equations defining $\mu ^{\pm }$ the
related Riemann-Hilbert problem is standard. We proceed through direct
computation of the difference of the two column vectors $\mu _1^{+}$ and $%
\mu _1^{-}$ and obtain the integral equation for this difference. It obeys
the same equation as the quantity $-\alpha ^{-}(k)\,k^{-n}\,\mu _2^{-}(k,n)$
(the quantity $\alpha ^{-}$ is defined below) and, based on the uniqueness
of the solution of such equations, we conclude
\begin{equation}
\mu _1^{+}(k,n)-\mu _1^{-}(k,n)=-\alpha ^{-}(k)\,k^{-n}\,\mu
_2^{-}(k,n),\quad |k|=1.  \label{RHmu1}
\end{equation}
The same approach is applied to $\mu _2^{\pm }$ and we get
\begin{equation}
\mu _2^{+}(k,n)-\mu _2^{-}(k,n)=\alpha ^{+}(k)\,k^n\,\mu _1^{+}(k,n),\quad
|k|=1  \label{RHmu2}
\end{equation}
where the {\em reflection coefficients} $\alpha ^{-}(k)$ and $\alpha ^{+}(k)$
are defined (still for $|k|=1$) as
\begin{equation}
\alpha ^{-}(k)=\sum_{-\infty }^{+\infty }k^ir(i)\mu _{11}^{-}(k,i),
\quad\alpha ^{+}(k)=\sum_{-\infty }^{+\infty }k^{-i}q(i)\mu _{22}^{+}(k,i).
\label{refco}
\end{equation}
For future use we define also
\begin{equation}
\beta ^{-}(k)=1-\sum_{-\infty }^{+\infty }q(i)\mu _{21}^{-}(k,i), \quad\beta
^{+}(k)=1-\sum_{-\infty }^{+\infty }r(i)\mu _{12}^{+}(k,i)  \label{transco}
\end{equation}
which are called the {\em transmission coefficients}. Note that, due to the
analytical properties of $\mu _{21}^{-}$ (resp. $\mu _{12}^{+}$), $\beta
^{-}(k)$ can be continued analytically in $|k|\ge 1$ (resp. $\beta ^{+}(k)$
in $|k|\le 1$). Actually the vectors $\mu _1^{-}$ and $\mu _2^{+}$ have
poles where the transmission coefficients $\beta ^{\pm }(k)$ have poles and
we derive from the integral equations (after multiplication by $k-k_j^{\pm }$
and limit $k\to k_j^{\pm }$)
\begin{eqnarray}
&&{\displaystyle\mathop{\mbox{Res }}_{k_j^{-}}}\{\mu _1^{-}(k,n)\}=\left( %
\matrix{0\cr1}\right) (k_j^{-})^{-n}C_j^{-}+\left( \matrix{-\displaystyle
\sum_{i=n+{1}}^{+\infty }q(i){\displaystyle\mathop{\mbox{Res }}_{k_j^{-}}}
\{\mu _{21}^{-}(k,i)\}\cr -\displaystyle\sum_{i=n+{1}}^{+\infty
}(k_j^{-})^{i-n}r(i){\displaystyle \mathop{\mbox{Res }}_{k_j^{-}}}\{\mu
_{11}^{-}(k,i)\}}\right) \\
&&{\displaystyle\mathop{\mbox{Res }}_{k_j^{+}}}\{\mu _2^{+}(k,n)\}=\left( %
\matrix{1\cr0}\right) (k_j^{+})^nC_j^{+}+\left(
\matrix{-\displaystyle \sum_{i=n+{1}}^{+\infty
}(k_j^{+})^{n-i}q(i){\displaystyle\mathop{ \mbox{Res
}}_{k_j^{+}}}\{\mu _{22}^{+}(k,i)\}\cr -\displaystyle\sum_{i=n+{1}}^{+\infty
}r(i){\displaystyle\mathop{\mbox{Res }} _{k_j^{+}}}\{\mu _{12}^{+}(k,i)\}}%
\right)
\end{eqnarray}
with the following definitions of the $C_j^{\pm }$'s
\begin{equation}
C_j^{-}=\sum_{i=-\infty }^{+\infty }(k_j^{-})^ir(i){\displaystyle %
\mathop{\mbox{Res }}_{k_j^{-}}}\{\mu _{11}^{-}(k,i)\}  \label{cj+}
\end{equation}
\begin{equation}
C_j^{+}=\sum_{i=-\infty }^{+\infty }(k_j^{+})^{-i}q(i){\displaystyle %
\mathop{\mbox{Res }}_{k_j^{+}}}\{\mu _{22}^{+}(k,i)\}  \label{cj-}
\end{equation}
which are called the {\em normalization coefficients}

Since the vectors $\mu _1^{+}$ and $\mu _2^{-}$ are holomorphic, we can
write down their integral equations evaluated respectively in $k_j^{+}$ and $%
k_j^{-}$ and compare them to the above integral equations for the residues.
We obtain
\begin{eqnarray}
&&{\displaystyle\mathop{\mbox{Res }}_{k_j^{-}}}\{\mu
_1^{-}(k,n)\}=(k_j^{-})^{-n}C_j^{-}\mu _2^{-}(k_j^{-},n)  \label{res-mu1} \\
&&{\displaystyle\mathop{\mbox{Res }}_{k_j^{+}}}\{\mu
_2^{+}(k,n)\}=(k_j^{+})^nC_j^{+}\mu _1^{+}(k_j^{+},n).  \label{res-mu2}
\end{eqnarray}

Finally the Riemann-Hilbert problem (\ref{RHmu1}) (\ref{RHmu2}) is completed
by the behaviors of the solution $\mu^\pm$ on the boundaries $|k|=0$ and $%
|k|=\infty$ of ${\cal D}$, which read
\begin{equation}
\mu_1^+(k,n)\quad\displaystyle{\mathop{\longrightarrow}_{k\to 0}} \hskip %
10pt\left(\matrix{1\cr0}\right),\quad \mu_2^-(k,n)\quad \displaystyle{\,%
\mathop{\longrightarrow}_{k\to\infty}} \quad\left( \matrix{0\cr1}\right).
\label{k-bounds}
\end{equation}
The vectorial Riemann-Hilbert problem (\ref{RHmu1}), (\ref{RHmu2}) with
singular points given by (\ref{res-mu1}), (\ref{res-mu2}) and the boundary
behaviors (\ref{k-bounds}) constitutes a closed problem which will be solved
in the next section.

The behaviors of $\mu _1^{-}(k,n)$ at large $k$ and of $\mu _2^{+}(k,n)$ at
small $k$ will be useful for the following and we define
\begin{equation}
\mu _{11}^{-}(k,n)\quad\displaystyle{\mathop{\longrightarrow}_{k\to \infty }}%
\quad f(n).
\end{equation}
Then
\begin{equation}
\mu _{21}^{-}(k,n)\quad\displaystyle{\mathop{\longrightarrow}_{k\to \infty }}%
\quad r(n)f(n)
\end{equation}
and one easily gets that $f(n)$ satisfies the integral equation
\begin{equation}
f(n)=1-\sum_{n+{1}}^{+\infty }q(i)r(i)f(i),
\end{equation}
the solution of which is
\begin{equation}
f(n)=\prod_{n+{1}}^{+\infty }[1-q(i)r(i)].  \label{f-de-n}
\end{equation}
The same computation holds for $\mu _2^{+}(k,n)$ and we have finally
\begin{eqnarray}
&&\mu ^{-}(k,n)\quad\displaystyle{\mathop{\longrightarrow}_{k\to \infty }}%
\left( \matrix{f(n) & 0\cr r(n)f(n) & 1}\right)  \label{muloin} \\
&&\mu ^{+}(k,n)\quad\displaystyle{\mathop{\longrightarrow}_{k\to 0}} \left( %
\matrix{ 1 & q(n)f(n) \cr 0 & f(n)}\right) .  \label{muproche}
\end{eqnarray}

It can be shown easily by using (\ref{RHmu1}) and (\ref{RHmu2}) that the
determinant of the matrix $\mu (k,n)$ is analytic in the whole domain ${\cal %
D}$. Hence it follows from the Liouville theorem and the boundary values (%
\ref{muloin}), (\ref{muproche}) that
\begin{equation}
\det \{\mu (k,n)\}=f(n).  \label{det-mu-rho}
\end{equation}
Note that, within the reduction $r=-\overline{q}$, $f(n)=\rho (n)^{-1}$ as
given by (\ref{S-prod-ener}).

\subsection{Asymptotic behaviors and unitarity relation}

By taking the limit at large $n$ directly on the integral equations, the
functions $\mu ^{\pm }$ obey for $|k|=1$
\begin{eqnarray}
&&\mu ^{-}(k,n)\quad\displaystyle{\mathop{\longrightarrow}_{n\to +\infty }}%
\quad\left( \matrix{1 & 0 \cr k^{-n}\alpha^-(k) & 1}\right)
\label{mumoins-nplus} \\
&&\mu ^{-}(k,n)\quad\displaystyle{\mathop{\longrightarrow}_{n\to -\infty }}%
\quad\left( \matrix{\beta^-(k) & -k^{n}\widehat{\alpha}^{\,-}(k)\cr 0 &
\widehat{\beta}^{\,-}(k) }\right)  \label{mumoins-nmoins} \\
&&\mu ^{+}(k,n)\quad\displaystyle{\mathop{\longrightarrow}_{n\to +\infty }}%
\quad\left( \matrix{1 & k^{n}\alpha^+(k) \cr 0& 1}\right)
\label{muplus-nplus} \\
&&\mu ^{+}(k,n)\quad\displaystyle{\mathop{\longrightarrow}_{n\to -\infty }}%
\quad\left( \matrix{\widehat{\beta}^+(k) & 0 \cr
-k^{-n}\widehat{\alpha}^+(k) & \beta^+(k) }\right)  \label{muplus-nmoins}
\end{eqnarray}
where the following {\em alternative scattering data} are defined as
\begin{eqnarray}
\widehat{\alpha}^{\,-}(k) &=&\sum_{-\infty }^{+\infty }k^{-i}q(i)\mu
_{22}^{-}(k,i),\quad\widehat{\alpha}^{+}(k)=\sum_{-\infty }^{+\infty
}k^ir(i)\mu _{11}^{+}(k,i)  \label{hatrefco} \\
\widehat{\beta}^{\,-}(k) &=&1-\sum_{-\infty }^{+\infty }r(i)\mu
_{12}^{-}(k,i), \quad\widehat{\beta}^{+}(k)=1-\sum_{-\infty }^{+\infty
}q(i)\mu _{21}^{+}(k,i).  \label{hattransco}
\end{eqnarray}
The quantities $\widehat{\alpha}^{\pm }$ are also called the reflection
coefficients {\em to the left} (referring to the limit $n\to -\infty $) when
$\alpha ^{\pm }$ are the reflection coefficients {\em to the right}.

It is easy to prove the following relations
\begin{eqnarray}
\widehat{\alpha}^{\,-} &=&{\frac{\alpha ^{+}\beta ^{-}}{1-\alpha ^{-}\alpha
^{+}}}, \quad\widehat{\alpha}^{+}={\frac{\alpha ^{-}\beta ^{+}}{1-\alpha
^{-}\alpha ^{+}}}  \label{alter-alpha} \\
\widehat{\beta}^{\,-} &=&{\frac{\beta ^{+}}{1-\alpha ^{-}\alpha ^{+}}},\quad
\widehat{\beta}^{+}={\,\frac{\beta ^{-}}{1-\alpha ^{-}\alpha ^{+}}}.
\label{alter-beta}
\end{eqnarray}
Indeed, by using the Riemann-Hilbert problems (\ref{RHmu1}) and (\ref{RHmu2}%
) for $\mu $ (still for $|k|=1$) we have
\begin{eqnarray}
\widehat{\alpha}^{\,-} &=&\sum_{-\infty }^{+\infty }k^{-i}q(i)\mu
_{22}^{+}(k,i)-\alpha ^{+}\sum_{-\infty }^{+\infty }q(i)\mu _{21}^{+}(k,i)
\nonumber \\
&=&\alpha ^{+}\left( 1-\sum_{-\infty }^{+\infty }q(i)\mu
_{21}^{+}(k,i)\right) =\alpha ^{+}\widehat{\beta}^{+} \\
\widehat{\beta}^{+} &=&1-\sum_{-\infty }^{+\infty }q(i)\mu
_{21}^{-}(k,i)+\alpha ^{-}\sum_{-\infty }^{+\infty }k^{-i}q(i)\mu
_{22}^{-}(k,i)  \nonumber \\
&=&\beta ^{-}+\alpha ^{-}\widehat{\alpha}^{\,-},
\end{eqnarray}
and so on for the other relations.

Now from (\ref{det-mu-n}) the determinant of $\mu(k,n)$ as $n\to-\infty$ can
be computed and, by using the behaviors of $\mu(k,n)$, it leads to the
relation
\begin{equation}
\beta^-\beta^+=(1-\alpha^-\alpha^+)\prod_{-\infty}^{+\infty}[1-r(i)q(i)]
\label{unitarity}
\end{equation}
which is called the {\em unitarity relation}.

\subsection{Reduction}

A {\em reduction} denotes a simple (possibly algebraic) relation between the
potentials (here $r(n)$ and $q(n)$) for which one can derive the counterpart
relations for the spectral data. In other word a reduction is a relation
which conserves the bijection between potentials and spectral data.

In the case
\begin{equation}
r(n)=-\overline{q}(n)\quad \Leftrightarrow \quad \overline{Q}(n)=\sigma
_2Q(n)\sigma _2,  \label{reduc-pot}
\end{equation}
it is easy to check that the function
\begin{equation}
\nu (k,n)=\sigma _2\overline{\mu }(1/\overline{k},n)\sigma _2,\qquad \sigma
_2=\left(
\begin{array}{cc}
0 & -i \\
i & 0
\end{array}
\right),  \label{reduc-nu}
\end{equation}
obeys the same equation as $\mu (k,n)$. To compare them it is enough to
consider their behaviors as $n\to \pm \infty $. Since
\begin{equation}
\sigma _2\overline{\mu }^{\,-}(1/\overline{k},n)\sigma _2\quad \displaystyle{%
\,\mathop{\longrightarrow}_{n\to +\infty }}\quad \left( \matrix{1 &
-k^{n}\overline{\alpha}^{\,-}(1/\overline{k}) \cr 0& 1}\right)
\end{equation}
we conclude that
\begin{eqnarray}
&&\sigma _2\overline{\mu }^{\,-}(1/\overline{k},n)\sigma _2=\mu ^{+}(k,n)
\label{reduc-mu} \\
&&\overline{\alpha }^{\,-}(1/\overline{k})=-\alpha ^{+}(k).
\label{reduc-alpha}
\end{eqnarray}
The same calculation at $n\to -\infty $ gives that
\begin{equation}
\overline{\beta }^{\,-}(1/\overline{k})=\beta ^{+}(k),  \label{reduc-beta}
\end{equation}
and also that the alternative scattering data obey similar relations.

For the discrete spectrum, the relation (\ref{reduc-mu}) implies
\begin{equation}
{\displaystyle\mathop{\mbox{Res }}_{k_j^{+}}}\{\mu _2^{+}(k,i)\}={\ %
\displaystyle \mathop{\mbox{Res}}_{k_j^{+}}}\left( \matrix{
-\overline{\mu}_{21}^{\,-}
(1/\overline{k},i)\cr\overline{\mu}_{11}^{\,-}(1/\overline{k},i)}\right).
\label{res-red1}
\end{equation}
Using the basic relations
\begin{equation}
{\displaystyle\mathop{\mbox{Res}}_{k_0}}\{f(1/\overline{k})\}=-(\overline{k}
_0)^2{\,\displaystyle\mathop{\mbox{Res}}_{1/\overline{k}_0}}\{f(k)\}, \quad{%
\ \displaystyle\mathop{\mbox{Res}}_{k_0}}\{\overline{g}(k)\}=\overline{%
\displaystyle \mathop{\mbox{Res}}_{k_0}}\{g(k)\},  \label{basic-res}
\end{equation}
the equation (\ref{res-red1}) becomes
\begin{equation}
{\displaystyle\mathop{\mbox{Res }}_{k_j^{+}}}\{\mu _2^{+}(k,i)\}=-(\overline{
k}_j^{+})^2\overline{\displaystyle\mathop{\mbox{Res}}_{1/\overline{k}_j^{+}}}
\left( \matrix{
-\overline{\mu}_{21}^{\,-}(k,i)\cr\overline{\mu}_{11}^{\,-}(k,i)}\right) .
\label{res-red2}
\end{equation}
This last equation holds if
\begin{equation}
N^{+}=N^{-},\quad k_j^{+}={\frac 1{\overline{k}_j^{\,-}}}  \label{reduc-kj}
\end{equation}
for which it reads
\begin{equation}
{\displaystyle\mathop{\mbox{Res }}_{k_j^{+}}}\{\mu _2^{+}(k,i)\}=-(\overline{
k}_j^{+})^2\overline{{\displaystyle\mathop{\mbox{Res}}_{k_j^{-}}}\left( %
\matrix{ -\overline{\mu}_{21}^{\,-}(k,i)\cr\overline{\mu}_{11}^{\,-}(k,i)}%
\right)}.  \label{res-red3}
\end{equation}
The above relation together with (\ref{reduc-mu}) and (\ref{reduc-kj})
implies then
\begin{equation}
{\frac{C_j^{+}}{k_j^{+}}}=
{\frac{\overline{C}_j^{\,-}}{\overline{k}_j^{\,-}}}.  \label{reduc-cj}
\end{equation}
In the case of the reduction (\ref{reduc-pot}) we will use the following
simplified notations (already used in sect. 2)
\begin{equation}
\alpha =\alpha ^{-},\quad \beta =\beta ^{-},\quad C_j=C_j^{-},\quad
k_j=k_j^{-},\quad N=N^{+}=N^{-}.  \label{not-alphabeta}
\end{equation}
In order to not over complicate this paper, we do not consider the other
reduction $r=\overline{q}$ for which similar results can be easily obtained,
but which corresponds to a spectral problem without discrete spectrum.

Last, it is useful for the following to rewrite the asymptotic boundary
behaviors (\ref{mumoins-nplus})-(\ref{muplus-nmoins}) within the reduction (and
with the above notations) and for $|\zeta|=1$ as:
\begin{equation}
\left( \matrix{\beta & \zeta^{n}\bar\alpha\gamma \cr 0 &
\bar\gamma }\right)
\displaystyle{\mathop{\longleftarrow}_{n\to -\infty }}\quad
\mu^{-}(\zeta,n)\quad\displaystyle{\mathop{\longrightarrow}_{n\to +\infty }}%
\quad\left( \matrix{1 & 0 \cr \zeta^{-n}\alpha & 1}\right)
\label{mu-n.behav} \end{equation}
\begin{equation}
\left( \matrix{\gamma & 0 \cr
-\zeta^{-n}\alpha\bar\gamma & \bar\beta }\right)
\quad\displaystyle{\mathop{\longleftarrow}_{n\to -\infty }}\quad
\mu^{+}(\zeta,n)\quad\displaystyle{\mathop{\longrightarrow}_{n\to +\infty }}%
\quad\left( \matrix{1 & -\zeta^{n}\bar\alpha \cr 0& 1}\right),
\label{mu+n.behav} \end{equation}
\begin{equation}
\gamma=\frac{\beta}{1+|\alpha|^2}.
 \end{equation}
Similarly, the unitarity relation (\ref{unitarity}) together with the definition
(\ref{f-de-n}) reads here
\begin{equation}
|\beta|^2=(1+|\alpha|^2)f(-\infty).
\label{unitar-reduc}
\end{equation}


\section{The inverse spectral problem}

The {\em inverse spectral problem} consists in reconstructing the potentials
$q(n)$ and $r(n)$ from the spectral data
\begin{eqnarray}
&&\alpha ^{\pm }(k),\quad \beta ^{\pm }(k),\quad |k|=1;  \nonumber \\
&&C_j^{\pm },\quad k_j^{\pm },\quad |k_j^{+}|<1,\quad j=1,..,N^{+},\quad
|k_j^{-}|>1,\quad j=1,..,N^{-}.  \label{spec-dat}
\end{eqnarray}
A simple way of doing this is to reformulate the analytical properties of
the matrix $\mu (k,n)$ in the domain ${\cal D}$ as a $\overline{\partial}$%
-problem.

\subsection{Inverse problem as a $\overline{\partial}$-problem}

Indeed, the set of fundamental relations (\ref{RHmu1}), (\ref{RHmu2}), (\ref
{res-mu1}) and (\ref{res-mu2}), which contain all the information about the
analytical properties of $\mu (k,n)$, can be summarized in the formula
\begin{equation}
{\frac \partial {\partial \overline{k}}}\,\mu (k,n)=\mu (k,n)R(k,n),\quad
k\in {\cal D},  \label{d-bar-mu}
\end{equation}
where the {\em spectral transform} contains all the information and reads
\begin{eqnarray}
&&R(k,n)=\left( \matrix{0&\alpha ^{+}(k)\delta ^{+}(k,1)\cr -\alpha
^{-}(k)\delta ^{-}(k,1)&0}\right) \left( \matrix{k^{-n}&0\cr 0&k^n}\right) -
\nonumber \\
&&-2i\pi \left( \matrix{0&\displaystyle\sum_{j=1}^{N^+}C_j^+
\delta(k-k_j^+)\cr \displaystyle\sum_{j=1}^{N^-}C_j^- \delta(k-k_j^-) & 0}%
\right) \left( \matrix{k^{-n} & 0 \cr 0 & k^{n} }\right) .
\label{d-bar-matrix}
\end{eqnarray}
The distributions $\delta ^{\pm }(k,1)$ are defined in the appendix and the
distribution $\delta (k)$ is normalized by requiring that $\int \!\!\!\int
dk\wedge d\overline{k}\,\delta (k)=1$. Using the method and tools described
in the appendix, and for the behaviors
\begin{equation}
\mu _1(k,n)\quad \displaystyle{\mathop{\longrightarrow}_{k\to 0}}\quad
\left( \matrix{ 1 \cr 0}\right) ,\quad \mu _2(k,n)\quad \displaystyle{%
\mathop{\longrightarrow}_{k\to \infty }}\quad \left( \matrix{ 0\cr
1}\right),  \label{d-bar-behav}
\end{equation}
we have the following solution of this $\overline{\partial }$-problem
\begin{equation}
\mu (k,n)=1+{\frac 1{2i\pi }}\int \!\!\!\!\int_{{\cal D}}{\frac{d\lambda
\wedge d\overline{\lambda }}{\lambda -k}}\,\mu (\lambda ,n)R(\lambda
,n)\left( \matrix{k/\lambda & 0 \cr 0 & 1 }\right) .  \label{sol-d-bar1}
\end{equation}
Due to the particular structure (\ref{d-bar-matrix}) of $R(k)$, the above
matrix valued equation has actually to be understood as two vectorial
integral equations, for $\mu _1^{+}(k,n)$ and for $\mu _2^{-}(k,n)$. As will
be seen hereafter, the knowledge of these two vectors is sufficient for
solving completely the problem.

\subsection{Reconstruction of the potentials}

The potentials are obtained from the asymptotic expansion of the Jost
solutions $\mu _2^{-}$ and $\mu _1^{+}$ of (\ref{spec-pb}) via the formulae
\begin{equation}
q(n{+}1)=-\mu _{12}^{-(-1)}(n),\quad r(n{+}1)=-\mu _{21}^{+(1)}(n),
\label{potentials}
\end{equation}
where $\mu _{12}^{-(-1)}(n)$ is the coefficient of $k^{-1}$ in the Laurent
expansion for $k\to \infty $ of $\mu _{12}^{-}(k,n)$, and $\mu
_{21}^{+(1)}(n)$ the coefficient of $k$ in the Taylor expansion for $k\to 0$
of $\mu _{21}^{+}(k,n)$. In particular from (\ref{sol-d-bar1}) we get
\begin{eqnarray}
r(n{+}1) &=&{\frac 1{2i\pi }}\oint_{{\cal C}}d\zeta \alpha ^{-}(\zeta )\zeta
^{-n-2}\mu _{22}^{-}(\zeta ,n)-\sum_{j=1}^{N^{-}}C_j^{-}(k_j^{-})^{-n-2}\
\mu _{22}^{-}(k_j^{-},n)  \label{r(n+1)} \\
q(n{+}1) &=&{\frac 1{2i\pi }}\oint_{{\cal C}}d\zeta \alpha ^{+}(\zeta )\zeta
^n\mu _{11}^{+}(\zeta ,n)+\sum_{j=1}^{N^{+}}C_j^{+}(k_j^{+})^n\,\mu
_{11}^{+}(k_j^{+},n)  \label{q(n+1)}
\end{eqnarray}
One could easily check that the potentials given by (\ref{q(n+1)}) and (\ref
{r(n+1)}) do obey the reduction (\ref{reduc-pot}) when $\mu (k,n)$ obeys (%
\ref{reduc-mu}), $\alpha (k)$ obeys (\ref{reduc-alpha}), $C_j^{\pm }$ obey (%
\ref{reduc-cj}) and $k_j^{\pm }$ obey (\ref{reduc-kj}).

Finally the transmission coefficients $\beta ^{-}(k)$ and $\beta ^{+}(k)$
are computed from their definitions (\ref{transco}), where the entries $\mu
_{21}^{-}(k,n)$ and $\mu _{12}^{+}(k,n)$ are obtained from the solution $\mu
_1^{+}$ and $\mu _2^{-}$ by using the explicit relations (\ref{RHmu1}), (\ref
{RHmu2}). Equivalently one can use the relations (\ref{alter-beta}) and the
behaviors (\ref{mumoins-nmoins}) and (\ref{muplus-nmoins}) to get ($|\zeta
|=1$)
\begin{eqnarray}
\beta ^{+}(\zeta ) &=&[1-\alpha ^{+}\alpha ^{-}]\,\displaystyle{\ %
\mathop{\mbox{lim}}_{n\to -\infty }}\,\mu _{22}^{-}(\zeta ,n)
\label{beta-m-calc} \\
\beta ^{-}(\zeta ) &=&[1-\alpha ^{+}\alpha ^{-}]\,\displaystyle{\ %
\mathop{\mbox{lim}}_{n\to -\infty }}\,\mu _{11}^{+}(\zeta ,n)
\label{beta-p-calc}
\end{eqnarray}

{\bf Remark.} From the other components of $\mu _1^{+}$ and $\mu _2^{-}$ we
obtain in ( \ref{spec-pb}) the following relations
\begin{equation}
\mu _{22}^{-(-1)}(n)=\sum_{n+{1}}^\infty r(i)q(i+1),\quad\mu
_{11}^{+(1)}(n)=\sum_{n+{1}}^\infty r(i+1)q(i).
\end{equation}


\section{The method of the $\overline{\partial}$-problem}

We have shown in the preceding sections that the spectral problem (\ref
{spec-pb}) can be mapped to the $\overline{\partial}$-problem (\ref{d-bar-mu}%
), namely
\begin{equation}
{\frac \partial {\partial \overline{k}}}\,\mu (k)=\mu (k)\,R(k), \quad k\in
{\cal D},  \label{d-bar}
\end{equation}
with the boundary behaviors (\ref{d-bar-behav}). The solution of such a
boundary value problem in the complex plane solves the Cauchy-Green integral
equation
\begin{equation}
\mu (k)=1+{\frac 1{2i\pi }}\int \!\!\!\!\int_{{\cal D}}{\frac{d\lambda
\wedge d\overline{\lambda }}{\lambda -k}}\,\mu (\lambda )R(\lambda )\left( %
\matrix{k/\lambda & 0 \cr 0 & 1 }\right) .  \label{sol-d-bar}
\end{equation}

The purpose of the following is to show that the above integral equation,
for the unknown $\mu $ and the datum $R$, can be taken as the {\em starting
tool}. More precisely we shall show how a parametric dependence of $R$ (on
an integer $n$ and on a real $t$) induces the spectral problem (\ref{spec-pb}%
) and a nonlinear evolution equation.

\subsection{The principal spectral problem}

We restrict this study to off-diagonal matrices $R(k)$ and consider the
integral equation (\ref{sol-d-bar}) as the given tool. If $R(k)$ depends now
on an external integer $n$, the solution $\mu (k,n)$ solves then the $%
\overline{\partial}$-problem (\ref{d-bar}) with the behaviors
\begin{eqnarray}
&&\mu (k,n)\quad\displaystyle{\mathop{\longrightarrow}_{k\to 0}} \quad\left( %
\matrix{ 1 & g(n)\cr 0 & f(n)}\right) +k\mu ^{(1)}(n)+\dots
\label{d-bar-behav-0} \\
&&\mu (k,n)\quad\displaystyle{\mathop{\longrightarrow}_{k\to \infty }}\hskip %
10pt\left( \matrix{f'(n) & 0\cr h(n) & 1}\right) +{\frac 1k}\mu
^{(-1)}(n)+\dots  \label{d-bar-behav-inf}
\end{eqnarray}
where the functions $f$, $f^{\prime }$, $g$ and $h$ have to be evaluated.
The determinant of $\mu (k,n)$ is analytic in ${\cal D}$ as indeed the
off-diagonal structure of $R(k,n)$ implies
\begin{equation}
{\frac \partial {\partial \overline{k}}}\det \{\mu (k,n)\}=0,
\label{d-bar-det}
\end{equation}
and from the above behavior the Liouville theorem implies
\begin{equation}
\det \{\mu (k,n)\}=f(n)=f^{\prime }(n).  \label{f-f'}
\end{equation}

We chose now the following explicit dependence of $R(k,n)$ on the discrete
variable $n$
\begin{equation}
R(k,n{+}1)=\Lambda (k)^{-1}R(k,n)\Lambda (k),  \label{R-fonc-n}
\end{equation}
with $\Lambda (k)$ defined in (\ref{Lambda}). The basic fundamental property
which allows us to derive from the choice (\ref{R-fonc-n}) a {\em difference}
equation for $\mu $ in the variable $n$ is the following
\begin{equation}
{\frac \partial {\partial \overline{k}}}\,H(k,n)=0,\quad H(k,n)=\mu (k,n{+}
1)\Lambda (k)^{-1}\mu (k,n)^{-1}\Lambda (k).  \label{fund-H-eq}
\end{equation}

The above function $H(k,n)$ can then be reconstructed from its behaviors on
the boundary of ${\cal D}$ ($k=\infty $ and $k=0$) which read from (\ref
{d-bar-behav-0}) and (\ref{d-bar-behav-inf})
\begin{eqnarray}
&&H(k,n)\quad\displaystyle{\mathop{\longrightarrow}_{k\to 0}}\quad{\ \,\frac
1{f(n)}}\left( \matrix{ f(n)-g(n{+}1)\mu_{21}^{(1)}(n) & g(n{+}1)\cr
-f(n{+}1)\mu_{21}^{(1)}(n) & f(n{+}1)}\right)  \label{H-behav-0} \\
&&H(k,n)\quad\displaystyle{\mathop{\longrightarrow}_{k\to \infty }} \hskip %
10pt{\frac 1{f(n)}}\left( \matrix{ f(n{+}1) & -f(n{+}1)\mu_{12}^{(-1)}(n)\cr
h(n{+}1) & f(n)-h(n{+}1)\mu_{12}^{(-1)}(n)} \right) .  \label{H-behav-inf}
\end{eqnarray}
Since $H(k,n)$ is analytic, these two behaviors are equal, which implies the
following four equations
\begin{eqnarray}
f(n{+}1) &=&f(n)-g(n{+}1)\mu _{21}^{(1)}(n)  \nonumber \\
f(n{+}1) &=&f(n)-h(n{+}1)\mu _{12}^{(-1)}(n)  \nonumber \\
g(n{+}1) &=&-f(n{+}1)\mu _{12}^{(-1)}(n)  \nonumber \\
h(n{+}1) &=&-f(n{+}1)\mu _{21}^{(1)}(n),  \label{def-H-behav}
\end{eqnarray}
which are solved by first {\em defining} the {\em potentials} as
\begin{equation}
q(n{+}1)=-\mu _{12}^{(-1)}(n),\quad r(n{+}1)=-\mu _{21}^{(1)}(n),
\label{pots}
\end{equation}
and hence
\begin{equation}
g(n)=q(n)f(n),\quad h(n)=r(n)f(n),  \label{g-h-f}
\end{equation}
with the recursion relation for $f(n)$
\begin{equation}
f(n)=f(n{+}1)[1-r(n{+}1)q(n{+}1)]  \label{recur-f}
\end{equation}
of which the solution is indeed given by (\ref{f-de-n}).

Finally, the solution of the $\overline{\partial}$-problem (\ref{fund-H-eq})
reads
\begin{equation}
H(k,n)={\frac 1{1-r(n{+}1)q(n{+}1)}}\left( \matrix{ 1 & q(n{+}1)\cr r(n{+}1)
& 1}\right)  \label{sol-H}
\end{equation}
which can be written with (\ref{fund-H-eq}) as the {\em discrete spectral
problem}
\begin{equation}
\mu (k,n{+}1)\Lambda (k)^{-1}\mu (k,n)^{-1}\Lambda (k)=[1-Q(n{+}1)]^{-1}
\end{equation}
where
\begin{equation}
Q(n)=\left( \matrix{ 0 & q(n)\cr r(n) & 0}\right) .  \label{Q}
\end{equation}
It is convenient for the following to define the quantity
\begin{equation}
U(n)=[1-Q(n{+}1)]^{-1}  \label{U}
\end{equation}
such that the equation for $\mu (k,n)$ reads
\begin{equation}
\mu (k,n{+}1)=U(n)\Lambda (k)^{-1}\mu (k,n)\Lambda (k)  \label{spec-pb-U}
\end{equation}
which is precisely the spectral problem (\ref{spec-pb}).

\subsection{Nonanalytic Dispersion Relations. A Theorem}

We consider now that $R(k,n)$ depends also on an external real $t$ and
address the problem of computing the expression of the time dependence of
the solution $\mu$ of (\ref{sol-d-bar}). The result can be stated as a
theorem.

{\bf Theorem.}
{\em When the spectral transform $R(k,n,t)$ evolves according to
\begin{equation}
R_t(k,n,t)=[R(k,n,t)\,,\,\Omega (k,t)]+M(k,n,t)  \label{evol-R}
\end{equation}
where
\begin{equation}
M(k,n{+}1,t)=\Lambda (k)^{-1}M(k,n,t)\Lambda (k),\quad[\Lambda (k)\,,\
\Omega (k,t)]=0,  \label{constr-evol-R}
\end{equation}
and where $\Omega (k,t)$ is the nonanalytic dispersion relation
\begin{equation}
\Omega (k,t)={\frac 1{2i\pi }}\int \!\!\!\!\int_{{\cal D}}{\frac{d\lambda
\wedge d\overline{\lambda }}{\lambda -k}}\,{\frac{\partial \Omega (\lambda
,t)}{\partial \overline{\lambda }}}\left( \matrix{k/\lambda & 0\cr 0 & 1}
\right)  \label{def-omega-sing}
\end{equation}
the potential $Q$ obeys the following evolution
\begin{equation}
Q_t(n{+}1,t)=\left[ \sigma _3\,,\,{\frac 1{2i\pi }}\int \!\!\!\!\int_{{\cal %
D }}{\frac{d\lambda \wedge d\overline{\lambda }}{2\lambda }}\,T(\lambda
,n,t)\right] ,  \label{evol-Q-sing}
\end{equation}
where}
\begin{equation}
T(k,n,t)=\Lambda (k)^{-1}\mu (k,n,t)\{M(k,n,t)-{\frac{\partial \Omega (k,t)}{
\partial \overline{k}}}\}\Lambda (k)\,\mu ^{-1}(k,n{+}1,t).  \label{T}
\end{equation}
This theorem is proved hereafter.

\underline{{\em The auxiliary spectral problem}}

Let us define the matrix
\begin{equation}
V(k,n,t)=\{\mu _t(k,n,t)-\mu (k,n,t)\Omega (k,t)\}\mu ^{-1}(k,n,t),
\label{V}
\end{equation}
and compute its $\overline{\partial}$-derivative which, from (\ref{d-bar-mu}%
), (\ref{evol-R}) and (\ref{constr-evol-R}), obeys
\begin{equation}
{\frac{\partial V(k,n,t)}{\partial \overline{k}}}=\mu (k,n,t)\{M(k,n,t)-{\
\frac{\partial \Omega (k,t)}{\partial \overline{k}}}\}\mu ^{-1}(k,n,t).
\label{dbar-eq-V}
\end{equation}

To solve the above $\overline{\partial}$-problem we need the behaviors of $%
V_1$ as $k\to 0$ and $V_2$ as $k\to \infty $. Since
\begin{eqnarray}
&&\mu (k,n,t)\quad\displaystyle{\mathop{\longrightarrow}_{k\to 0}} \hskip %
10pt\left( \matrix{ 1 & q(n,t)f(n,t)\cr 0 & f(n,t)}\right) +k\mu
^{(1)}(n,t)+...  \label{mu-behav-0} \\
&&\mu (k,n,t)\quad\displaystyle{\mathop{\longrightarrow}_{k\to \infty }}
\quad\left( \matrix{f(n,t) & 0\cr r(n,t)f(n,t) & 1}\right) +{\frac 1k} \mu
^{(-1)}(n,t)+...  \label{mu-behave-inf}
\end{eqnarray}
it is easy to obtain, thanks also to the choice (\ref{def-omega-sing}) (it
would not be so in the case of a regular dispersion relation)
\begin{equation}
V_1(k,n,t)\quad\displaystyle{\mathop{\longrightarrow}_{k\to 0}} \quad\left( %
\matrix{ 0\cr 0 }\right) ,\quad V_2(k,n,t)\quad \displaystyle{%
\mathop{\longrightarrow}_{k\to \infty }}\quad\left( \matrix{0 \cr 0}\right) .
\label{V-behav}
\end{equation}
Consequently the solution reads
\begin{equation}
V(k,n,t)={\frac 1{2i\pi }}\int \!\!\!\!\int_{{\cal D}}{\frac{d\lambda \wedge
d\overline{\lambda }}{\lambda -k}}\,S(\lambda ,n,t)\left( \matrix{k/\lambda
& 0\cr 0 & 1}\right) .  \label{sol-V}
\end{equation}
where we have defined
\begin{equation}
S(k,n,t)=\mu (k,n,t)\{M(k,n,t)-{\frac{\partial \Omega (k,t)}{\partial
\overline{k}}}\}\mu ^{-1}(k,n,t).  \label{S}
\end{equation}

Now, with the above value of $V$, the definition (\ref{V}) can be written as
the {\em auxiliary spectral problem}
\begin{equation}
\mu_t(k,n,t)=V(k,n,t)\mu(k,n,t)+\mu(k,n,t)\Omega(k,t),  \label{aux-spec-pb}
\end{equation}

\underline{{\em Fundamental property of $V(k,n,t)$}}

For simplicity of notations, we omit from now on the variable $t$. By direct
computation the matrix $S$ obeys the equation
\begin{equation}
S(k,n{+}1)U(n)=U(n)\Lambda (k)^{-1}S(k,n)\Lambda (k).  \label{relat-S}
\end{equation}
The next step consists in seeking an analogous property for $V(k,n)$, by
computing the quantities $U(n)^{-1}V(k,n{+}1)$ on one side and $\Lambda
(k)^{-1}V(k,n)\Lambda (k)U(n)^{-1}$ on the other side. Using
\begin{eqnarray*}
&&\Lambda (\lambda )^{-1}S(\lambda ,n)\Lambda (\lambda )U(n)^{-1}\left( %
\matrix{k/\lambda & 0\cr 0 & 1}\right) -\Lambda (k)^{-1}S(\lambda ,n)\left( %
\matrix{k/\lambda & 0\cr 0 & 1}\right) \Lambda (k)U(n)^{-1} \\
&&\quad ={\frac{\lambda -k}\lambda }\left( \matrix{0&-q(n{+}1)s_{11}(\lambda
,n)+\lambda s_{12}(\lambda ,n)\cr r(n{+}1)s_{22}(\lambda ,n)-(1/\lambda
)s_{21}(\lambda ,n)&0}\right)
\end{eqnarray*}
we obtain finally the required property of $V(k,n)$
\begin{eqnarray}
&&U(n)^{-1}V(k,n{+}1)-\Lambda (k)^{-1}V(k,n)\Lambda (k)U(n)^{-1}=P(n)
\label{relat-V} \\
&&P(n)={\frac 1{2i\pi }}\int \!\!\!\!\int_{{\cal D}}{\frac{d\lambda \wedge d
\overline{\lambda }}{2\lambda }}\,[\sigma _3\,,\,T(\lambda ,n)]  \label{P} \\
&&T(k,n)=\Lambda (k)^{-1}S(k,n)\Lambda (k)U(n)^{-1}=U(n)^{-1}S(k,n{+}1).
\end{eqnarray}
We have used here above (\ref{relat-S}) to rewrite $T$ in a more convenient
form, and finally from the definition (\ref{S}) of $S(k,n)$ and the spectral
problem (\ref{spec-pb}) it reads
\begin{equation}
T(k,n)=\Lambda (k)^{-1}\mu (k,n)\{M(k,n)-{\frac{\partial \Omega (k)}{
\partial \overline{k}}}\}\Lambda (k)\,\mu ^{-1}(k,n{+}1).
\end{equation}
Note that the matrix $T(k,n)$ obeys a property similar to $S(k,n)$ since it
can be checked directly that
\begin{equation}
U(n-1)T(k,n-1)=\Lambda (k)T(k,n)U(n)\Lambda (k)^{-1}.  \label{relat-T}
\end{equation}

\underline{{\em The evolution equation}}

The nonlinear evolution of $Q(n,t)$ is now obtained in the usual way by
requiring the compatibility between (\ref{spec-pb-U}) and (\ref{aux-spec-pb}%
) which reads
\begin{eqnarray}
{\frac \partial {\partial t}}\mu (k,n{+}1,t) &=&{\frac \partial {\partial t}}
\{U(n)\Lambda (k)^{-1}\mu (k,n)\Lambda (k)\}  \nonumber \\
&=&V(k,n{+}1,t)\mu (k,n{+}1,t)+\mu (k,n{+}1,t)\Omega (k,t).
\end{eqnarray}
By means of (\ref{relat-V}) it is then easy to obtain the equation
\begin{equation}
U_t(n)=U(n)P(n)U(n),
\end{equation}
which readily gives the evolution (\ref{evol-Q-sing}) since $U(n)=[1-Q(n{+}
1)]^{-1}$. This ends the proof of the theorem.

\subsection{ Reduction}

If we consider the reduction $r(n)=-\overline{q}(n)$ the matrices
\begin{eqnarray}
&&\Omega (k,t)=\left( \matrix{\omega_1(k,t) & 0\cr 0 &\omega_2(k,t)}\right)
\label{O-def} \\
&&M(k,n,t)=\left( \matrix{0 & m_2(k,t)\cr m_1(k,t) & 0}\right) \left( %
\matrix{k^{-n} & 0 \cr 0 & k^{n} }\right) .  \label{M-def}
\end{eqnarray}
must be compatible with the preservation of the structure of $R(n)$ in the
time evolution equation (\ref{evol-R}). We choose, therefore, $\omega _1$
and $\omega _2$ analytic inside and outside the unite circle with limit
values on the two sides of the circle satisfying the symmetry properties ($%
\zeta =e^{i\theta }$)
\begin{eqnarray}
\omega _1^{-}(\zeta ,t) &=&\overline{\omega _2^{+}}(\zeta ,t)
\label{Omega-reduc} \\
\omega _2^{-}(\zeta ,t) &=&\overline{\omega _1^{+}}(\zeta ,t)  \nonumber
\end{eqnarray}
and we choose $m_1$ and $m_2$ as
\begin{eqnarray}
m_1^{}(k,t) &=&m^{-}(k,t)\delta ^{-}(k,1)\equiv m(\zeta ,t)\delta ^{-}(k,1)
\label{M-reduc} \\
m_2^{}(k,t) &=&m^{+}(k,t)\delta ^{+}(k,1)\equiv \overline{m}(\zeta ,t)\delta
^{+}(k,1)  \nonumber
\end{eqnarray}
where
\[
\zeta =\frac k{|k|}
\]
and $m(\zeta ,t)$ is a given {\em function} defined on the circle $|k|=1$.
Note that the discontinuity of $\omega _1$%
\begin{equation}
p(\zeta ,t)\equiv \omega _1^{+}(\zeta ,t)-\omega _1^{-}(\zeta ,t)
\end{equation}
is related to the discontinuity of $\omega _2$ by the formula
\begin{equation}
\omega _2^{+}(\zeta ,t)-\omega _2^{-}(\zeta ,t)=-\overline{p}(\zeta ,t).
\end{equation}
The analytic properties of $\omega _1$ and $\omega _2$ are summarized by the
formulae ($\zeta =k/|k|$)
\begin{eqnarray}
\frac{\partial \omega _1}{\partial \overline{k}} &=&p(\zeta )\delta (k,1) \\
\frac{\partial \omega _2}{\partial \overline{k}} &=&-\overline{p}(\zeta
)\delta (k,1)
\end{eqnarray}
where the distribution $\delta (k,1)$ is defined in the appendix. Requiring
that $\omega _1\rightarrow 0$ for $k\rightarrow 0$ and $\omega _2\rightarrow
0$ for $k\rightarrow \infty $, $\Omega (k)$ is defined by the following
Cauchy-Green formula
\begin{equation}
\Omega (k)=\frac 1{2\pi i}\oint_{{\cal C}}\frac{d\zeta }{\zeta -k}\left(
\begin{array}{cc}
p(\zeta ,t) & 0 \\
0 & -\overline{p}(\zeta ,t)
\end{array}
\right) \left(
\begin{array}{cc}
k/\zeta  & 0 \\
0 & 1
\end{array}
\right) .  \label{Omega}
\end{equation}
It results that
\begin{equation}
\omega _1(k)=\overline{\omega _2}(1/\overline{k})
\end{equation}
in agreement with the conditions on the boundaries (\ref{Omega-reduc}).

It can be shown that with the choices indicated in (\ref{Omega-reduc}) and (%
\ref{M-reduc}) the evolution equation (\ref{evol-Q-sing}) is compatible with
the reduction $r=-\overline{q}$. This evolution reads
\begin{equation}
q_t(n{+}1,t)={\frac 1{2\pi }}{\frac 1{f(n{+}1,t)}}\int_{-\pi }^{+\pi
}d\theta \,\gamma (\theta ,n,t),  \label{evol-q-reduc}
\end{equation}
where
\begin{eqnarray}
&&\gamma (\theta ,n)=-\omega _1^{-}(\zeta )\mu _{11}^{-}(\zeta ,n)\mu
_{12}^{-}(\zeta ,n{+}1)+\omega _1^{+}(\zeta )\mu _{11}^{+}(\zeta ,n)\mu
_{12}^{+}(\zeta ,n{+}1)  \nonumber \\
&&+\omega _2^{-}(\zeta )\zeta \mu _{12}^{-}(\zeta ,n)\mu _{11}^{-}(\zeta ,n{+%
}1)-\omega _2^{+}(\zeta )\zeta \mu _{12}^{+}(\zeta ,n)\mu _{11}^{+}(\zeta ,n{%
+}1)  \nonumber \\
&&+\alpha (\zeta )(\omega _1^{-}(\zeta )-\omega _2^{-}(\zeta ))\zeta
^{-n}\mu _{12}^{-}(\zeta ,n)\mu _{12}^{-}(\zeta ,n{+}1)  \nonumber \\
&&+\overline{\alpha }(\zeta )(\omega _1^{+}(\zeta )-\omega _2^{+}(\zeta
))\zeta ^{n+{1}}\mu _{11}^{+}(\zeta ,n)\mu _{11}^{+}(\zeta ,n{+}1)  \nonumber
\\
&&-m(\zeta )\zeta ^{-n}\mu _{12}^{-}(\zeta ,n)\mu _{12}^{-}(\zeta ,n{+}1)+%
\overline{m}(\zeta )\zeta ^{n{+}1}\mu _{11}^{+}(\zeta ,n)\mu _{11}^{+}(\zeta
,n{+}1).  \label{def-evol-q}
\end{eqnarray}
with $\zeta =e^{i\theta }$ (remember that $\alpha \equiv \alpha ^{-}$ and $%
\alpha ^{+}(\zeta )=-\overline{\alpha }(\zeta )$). Note that in computing $%
T(\lambda ,n)$ in (\ref{evol-Q-sing}) the term containing $\Omega (k,t)$
must be written as follows
\begin{eqnarray}
&&\Lambda ^{-1}\mu (n)\frac{\partial \Omega }{\partial \overline{k}}\Lambda
\mu ^{-1}(n{+}1)=  \nonumber \\
&&\frac \partial {\partial \overline{k}}\left( \Lambda ^{-1}\mu (n)\Omega
\Lambda \mu ^{-1}(n{+}1)\right) -\Lambda ^{-1}\mu (n)\left[ R(n),\Omega
\right] \Lambda \mu ^{-1}(n{+}1)  \label{dbarOmega}
\end{eqnarray}
which is a well defined local formulation of a $\overline{\partial }$%
-problem for a sectionally holomorphic function.

It is convenient, by using equations (\ref{RHmu1}) and (\ref{RHmu2}), to
rewrite (\ref{def-evol-q}) in terms of $\mu _1^{+}$ and $\mu _2^{-}$%
\begin{eqnarray}
&&\gamma (\theta ,n)=p(\zeta )\mu _{11}^{+}(\zeta ,n)\mu _{12}^{-}(\zeta ,n{+%
}1)+\overline{p}(\zeta )\zeta \mu _{12}^{-}(\zeta ,n)\mu _{11}^{+}(\zeta ,n{+%
}1)  \nonumber \\
&&-m(\zeta )\zeta ^{-n}\mu _{12}^{-}(\zeta ,n)\mu _{12}^{-}(\zeta ,n{+}1)+%
\overline{m}(\zeta )\zeta ^{n+{1}}\mu _{11}^{+}(\zeta ,n)\mu _{11}^{+}(\zeta
,n{+}1)  \label{evol-q-mu+}
\end{eqnarray}
which shows explicitly that the evolution equation depends only on the
discontinuity of $\Omega (k)$ on the unit circle.


\section{Integrable Discrete Initial-Boundary Value Problem}

By using the tools previously developed we prove now that the nonlinear
system (\ref{basic-sys}) is integrable when it is related to the initial
boundary value (\ref{data-bound}). We rewrite hereafter this system in the
variable $\zeta =e^{i\theta }$ and with the relation (\ref{S-prod-ener}) as
\begin{eqnarray}
q_t(n,t)\prod_{i=n{+}1}^\infty (1+|q(i,t)|^2) &=&\frac 1{2\pi }\int_{-\pi
}^{+\pi }d\theta e^{in\theta }(A_1*A_2)(\zeta ,n,t)  \label{evol-q-phys} \\
A_1(\zeta ,n,t)-A_1(\zeta ,n{-}1,t) &=&\zeta ^{-n}q(n,t)A_2(\zeta ,n,t) \\
A_2(\zeta ,n,t)-A_2(\zeta ,n{-}1,t) &=&-\zeta ^n\overline{q}(n,t)A_1(\zeta
,n,t)  \label{zs-phys}
\end{eqnarray}
where the interaction term is defined as
\begin{equation}
(A_1*A_2)(\zeta ,n,t)=g(\theta ,t)A_1(\zeta ,n{-}1,t)\overline{A}_2(\zeta
,n,t)+\overline{g}(\theta ,t)A_1(\zeta ,n,t)\overline{A}_2(\zeta ,n{-}1,t).
\label{by-prod}
\end{equation}

{\bf Theorem.}
{\em With the datum of the initial value $q(n,0)$ and the following
arbitrary boundary values as $n\to +\infty $
\begin{equation}
A_1(\zeta ,n,t)\to I_1(\zeta ,t),\quad A_2(\zeta ,n,t)\to I_2(\zeta ,t).
\label{phys-bounds}
\end{equation}
the above system is solvable by the spectral transform method.}

\subsection{Proof of integrability}

The proof is performed by showing that the evolution (\ref{evol-q-phys}) can
actually be written under the form (\ref{evol-q-reduc}). This then gives a
unique definition of the functions $m(k)$ and $\omega (k)+\overline{\omega }
(k)$ for which the two equations (\ref{evol-q-reduc}) and (\ref{evol-q-phys}%
) are {\em identical}. Hence the evolution (\ref{evol-R}) of the spectral
transform $R(k,t)$ is uniquely given via (\ref{Omega-reduc}) and (\ref
{M-reduc}).

The first useful property is that {\em the following 5 vectors}
\begin{equation}
\left(\matrix{A_1(\zeta,n)\cr \zeta^{-n}A_2(\zeta,n)}\right),\quad \left( %
\matrix{\mu_{11}^\pm(\zeta,n)\cr \mu_{21}^\pm(\zeta,n)}\right),\quad
\zeta^{-n}\left(\matrix{\mu_{12}^\pm(\zeta,n)\cr \mu_{22}^\pm(\zeta,n)}
\right),  \label{base-vec-zs}
\end{equation}
{\em solve the equation} (\ref{zs-phys}). Then, by comparison of their
asymptotic behaviors as $n\to+\infty$ given in (\ref{phys-bounds}) and in (%
\ref{muplus-nplus})-(\ref{mumoins-nmoins}), we get
\begin{equation}
\left(\matrix{A_1(\zeta,n)\cr \zeta^{-n}A_2(\zeta,n)}\right)=
I_1(\zeta)\left(\matrix{\mu_{11}^+(\zeta,n)\cr \mu_{21}^+(\zeta,n)}\right)+
I_2(\zeta)\zeta^{-n}\left(\matrix{\mu_{12}^-(\zeta,n)\cr \mu_{22}^-(\zeta,n)}
\right).  \label{exp-mu-base}
\end{equation}

Next, to compute the product $(A_1*A_2)(\zeta ,n)$ we make use of the
Riemann-Hilbert relations
\begin{eqnarray}
\mu _1^{-}(\zeta ,n)-\mu _1^{+}(\zeta ,n) &=&\alpha ^{-}(\zeta )\zeta
^{-n}\mu _2^{-}(\zeta ,n) \\
\mu _2^{-}(\zeta ,n)-\mu _2^{+}(\zeta ,n) &=&-\alpha ^{+}(\zeta )\zeta ^n\
\mu _1^{+}(\zeta ,n)
\end{eqnarray}
and rewrite it in terms only of $\mu _1^{+}$ and $\mu _2^{-}$%
\begin{eqnarray}
\zeta ^{n+1}(A_1*A_2)(\zeta ,n{+}1) &=&-\left[ g|I_1|^2-\overline{g}%
|I_2|^2\right] \mu _{11}^{+}(n)\mu _{12}^{-}(n{+}1)  \nonumber \\
&&-\left[ \overline{g}|I_1|^2-g|I_2|^2\right] \zeta \mu _{12}^{-}(n)\mu
_{11}^{+}(n{+}1)  \nonumber \\
&&-(g+\overline{g})\overline{I_1}I_2\zeta ^{-n}\mu _{12}^{-}(n)\mu
_{12}^{-}(n{+}1)  \nonumber \\
&&+(g+\overline{g})I_1\overline{I_2}\zeta ^{n{+}1}\mu _{11}^{+}(n)\mu
_{11}^{+}(n{+}1)
\end{eqnarray}

Then, thanks to the expression (\ref{f-de-n}), the two equations (\ref
{evol-q-reduc}) and (\ref{evol-q-phys}) are {\em identical} if and only if
\begin{eqnarray}
&&p(\zeta ,t)=-g(\theta ,t)|I_1(\theta ,t)|^2+\overline{g}(\theta
,t)|I_2(\theta ,t)|^2  \label{p-de-k} \\
&&m(\zeta ,t)=(g(\theta ,t)+\overline{g}(\theta ,t))\overline{I}_1(\theta
,t)I_2(\theta ,t).  \label{m-de-k}
\end{eqnarray}

Finally the theorem is proved and it remains to compute the evolution of the
spectral transform.

\subsection{Evolution of the spectral transform}

\underline{\em Evolution of $\alpha(\zeta,t)$}

The time evolution of $R(k,n,t)$ is given by (\ref{evol-R}) with $M$ and $%
\Omega $ defined in (\ref{M-reduc}) and in (\ref{Omega}). Taking into
account the structure (\ref{d-bar-matrix}) of $R(k,n,t)$ we have
\begin{eqnarray}
&&\partial _t\alpha (\zeta ,t)=\left[ \omega _1^{-}(\zeta ,t)-\omega
_2^{-}(\zeta ,t)\right] \alpha (\zeta ,t)-m(\zeta ,t)
\label{evol-alpha}\\
&&\partial _tk_j=0,\quad \partial _tC_j(t)=\left[ \omega _1(k_j,t)-\omega
_2(k_j,t)\right] C_j(t)\label{evol-Cj}
\end{eqnarray}
where from (\ref{Omega}) and the Sokhotski--Plemelj formula we have

\begin{eqnarray}
\omega _1^{-}(\zeta ,t)-\omega _2^{-}(\zeta ,t) &=&-\frac 12p(\zeta
,t)-\frac 12\overline{p}(\zeta ,t)  \nonumber \\
&&+\frac 1{2\pi i}P\oint_{{\cal C}}\frac{\,d\zeta ^{\prime }}{\zeta ^{\prime
}-\zeta }\,p(\zeta ^{\prime },t)\,\frac \zeta {\zeta ^{\prime }} \\
&&+\frac 1{2\pi i}P\oint_{{\cal C}}\frac{\,d\zeta ^{\prime }}{\zeta ^{\prime
}-\zeta }\,\overline{p}(\zeta ^{\prime },t) \\
\omega _1(k_j,t)-\omega _2(k_j,t) &=&\frac 1{2\pi i}\oint_{{\cal C}}\frac{%
\,d\zeta ^{\prime }}{\zeta ^{\prime }-k_j}\,p(\zeta ^{\prime },t)\,\frac{k_j%
}{\zeta ^{\prime }} \\
&&+\frac 1{2\pi i}\oint_{{\cal C}}\frac{\,d\zeta ^{\prime }}{\zeta ^{\prime
}-k_j}\,\overline{p}(\zeta ^{\prime },t).
\end{eqnarray}
and the functions $p(\zeta ,t)$ and $m(\zeta ,t)$ are given in (\ref{p-de-k}%
) and (\ref{m-de-k}).

As a result the evolution equation of $\alpha $ and $C_j$ can be written
\begin{eqnarray}
\partial _t\alpha &=&\alpha \frac{g+\overline{g}}2\left( |I_1(\theta
,t)|^2-|I_2(\theta ,t)|^2\right) -(g+\overline{g})\overline{I_1}I_2 \\
&&-\alpha \frac 1{2\pi i}P\oint_{{\cal C}}\frac{\,d\zeta ^{\prime }}{\zeta
^{\prime }-\zeta }(g+\overline{g})(|I_1|^2-|I_2|^2) \\
&&+\alpha \frac 1{2\pi i}\oint_{{\cal C}}\frac{d\zeta ^{\prime }}{\zeta
^{\prime }}\left( g|I_1|^2-\overline{g}|I_2|^2\right) , \\
\partial _tC_j(t) &=&-C_j(t)\frac 1{2\pi i}P\oint_{{\cal C}}\frac{\,d\zeta
^{\prime }}{\zeta ^{\prime }-k_j}(g+\overline{g})(|I_1|^2-|I_2|^2) \\
&&+C_j(t)\frac 1{2\pi i}\oint_{{\cal C}}\frac{d\zeta ^{\prime }}{\zeta
^{\prime }}\left( g|I_1|^2-\overline{g}|I_2|^2\right) .
\end{eqnarray}

\underline{\em Evolution of $\beta(\zeta,t)$}

The definition (\ref{alter-beta}) allows to obtain readily
\begin{equation}\label{RH-beta}
\frac{\widehat{\beta }_t^{+}}{\widehat{\beta }^{+}}-\frac{\beta _t}\beta =-%
\frac{\left( |\alpha |^2\right) _t}{1+|\alpha |^2},
\end{equation}
which actually can be understood as a Riemann-Hilbert problem on the
unit circle. Its solution reads
$$
|k|>1\quad:\quad
\frac{\widehat{\beta }_t^{+}}{\widehat{\beta }^{+}}
=\frac \partial {\partial t}\left( \frac{|\beta |^2}{1+|\alpha |^2}%
\right) \frac{1+|\alpha |^2}{|\beta |^2}  $$
\begin{equation}
-\frac 1{2\pi i}\oint_{{\cal C}}\frac{d\zeta^{\prime} }{\zeta^{\prime} -k}\frac{(|\alpha
(\zeta^{\prime} )|^2)_t}{1+|\alpha (\zeta^{\prime} )|^2} ,
\end{equation}
\begin{equation}
|k|<1\quad:\quad
\frac{\beta _t}\beta=
-\frac 1{2\pi i}\oint_{{\cal C}}\frac{d\zeta^{\prime} }{\zeta^{\prime}
-k}\,\frac k{\zeta^{\prime}}\
\frac{(|\alpha (\zeta^{\prime} )|^2)_t}{1+|\alpha (\zeta^{\prime} )|^2}.
\end{equation}
Hence, writing the above equation for $k=\zeta(1-0)$, we get the evolution
(\ref{S-evol-beta}).

\subsection{Evolution of the spectral transform from the Lax pair}

For completeness, we rederive hereafter the preceding formula (evolution of
$\alpha$ and $\beta$, in the absence of bound states for simplicity), by using
the traditional approach for which the starting tool is the Lax pair
(\ref{spec-pb}) (\ref{aux-spec-pb}). The method consists simply in evaluating
the asymptotic boundary values as $n\to\pm\infty$ on the auxiliary spectral
problem (\ref{aux-spec-pb}), in which (forget for a while the $(n,t)$-dependence)
\begin{equation}
V(k) =\frac 1{2i\pi }\int \!\!\!\!\int_{{\cal D}}{\frac{d\lambda
\wedge d\overline{\lambda }}{\lambda -k}}\mu (\lambda)
\left( M(\lambda)-\frac{\partial \Omega (\lambda)}{\partial
\overline{\lambda }}\right)\mu ^{-1}(\lambda)\left(
\begin{array}{cc}
k/\lambda & 0 \\
0 & 1
\end{array}
\right) .  \label{V-spec-pb}
\end{equation}

By using the identity (\ref{dbarOmega}), the equation (\ref{aux-spec-pb})
can be more conveniently written as
\begin{equation}
\mu_t(k,n,t)=X(k,n,t)\mu(k,n,t),
\label{simpl-sp}\end{equation}
where
\begin{equation}\label{X-gener}
X(k) =\frac 1{2i\pi }\int \!\!\!\!\int_{{\cal D}}{\frac{d\lambda
\wedge d\overline{\lambda }}{\lambda -k}}\mu (\lambda ,n,t)
\left( M(\lambda)+[R(\lambda),\Omega(\lambda)]\right)
\mu ^{-1}(\lambda)\left(\begin{array}{cc}
k/\lambda & 0 \\
0 & 1
\end{array} \right) .
\end{equation}
By inserting in the above equation the explicit forms of $R(k,n,t)$
given in (\ref{d-bar-matrix}) with (\ref{reduc-alpha}) and no bound states,
of $\Omega(k,t)$ given in  (\ref{Omega}), and of $M(k,n,t)$ given in
(\ref{M-def}) with (\ref{M-reduc}), we get finally
\begin{equation}\label{X-reduc}
X(k,n,t)=\frac 1{2i\pi }\oint\frac{d\zeta^{\prime}}{\zeta^{\prime}-k}\
\frac{1}{f(n)}\ \chi(\zeta^{\prime},n,t)
\left(\begin{array}{cc} k/\zeta^{\prime} & 0 \\ 0 & 1 \end{array} \right),
\end{equation}
with the following definition
\begin{eqnarray}\label{chi}
\chi(\zeta)&=&\zeta^n[\bar m+(\omega_1^+-\omega_2^+)\bar\alpha]
\left(\matrix{ -\mu_{11}^+\mu_{21}^+ & (\mu_{11}^+)^2 \cr
                -(\mu_{21}^+)^2      & \mu_{11}^+\mu_{21}^+}\right)+
\nonumber\\
&&\zeta^{-n}[ m-(\omega_1^--\omega_2^-)\alpha]
\left(\matrix{ \mu_{12}^-\mu_{22}^- & -(\mu_{12}^-)^2 \cr
                (\mu_{22}^-)^2      & -\mu_{12}^-\mu_{22}^-}\right)
\end{eqnarray}

The main tool is now the asymptotic boundary behaviors
(\ref{mu-n.behav}) and (\ref{mu+n.behav}) of $\mu^\pm$
which allows to obtain, by taking the limit as $n\to+\infty$ of
$\zeta^n\partial_t\mu^-_{21}(\zeta,n,t)$, the relations
\begin{eqnarray*}
\alpha _t(k) &=&-\frac 12\left[ m-(\omega _1^{-}-\omega _2^{-})\alpha
\right] (k)+ \\
&&\lim_{n\rightarrow \infty }\frac 1{2\pi i}P\oint \frac{d\zeta }{\zeta -k}%
\left( \frac k\zeta \right) ^{n+1}\left[ m-(\omega _1^{-}-\omega
_2^{-})\alpha \right] (\zeta ) \\
0 &=&-\frac 12\left[ \overline{m}+(\omega _1^{+}-\omega _2^{+})\overline{%
\alpha }\right] (k)+ \\
&&\lim_{n\rightarrow \infty }\frac 1{2\pi i}P\oint \frac{d\zeta }{\zeta -k}%
\left( \frac \zeta k\right) ^n\left[ \overline{m}+(\omega _1^{+}-\omega
_2^{+})\overline{\alpha }\right] (\zeta ).
\end{eqnarray*}
Consequently,
with the formula (see appendix)
\begin{equation}
\lim_{n\rightarrow \infty }\frac 1{2\pi i}P\oint \frac{d\zeta }{\zeta -k}%
\left( \frac k\zeta \right) ^n\Phi (\zeta )=-\frac 12\Phi (k),\qquad |k|=1,
\label{lim-distr}
\end{equation}
the preceding relations result precisely in the required evolution
(\ref{evol-alpha}).

In the same way, by taking the limit as $n\to-\infty$ of
$\partial_t\mu^-_{11}(\zeta,n,t)$, we obtain readily the required
evolution (\ref{S-evol-beta}) of the transmission coefficient $\beta(\zeta,t)$
(note that there, one should use also the unitarity relation
(\ref{unitar-reduc})).

\subsection{ Time evolution of $f(n)$}

It could be useful to have also explicitly the time evolution of
the quantity $f(n)$.From (\ref{det-mu-rho}) we have
\begin{equation}
f_t(n)=f(n)\,\mbox{\rm tr}\left\{ \mu _t(k,n)\mu ^{-1}(k,n)\right\}
\end{equation}
and then using the auxiliary spectral problem (\ref{simpl-sp})
\begin{equation}
f_t(n)=f(n)\ \mbox{\rm tr}\left\{ X(k,n)\right\} .
\end{equation}
From the expression (\ref{X-reduc})
we obtain that the trace of $X(k,n)$ is $k$-independent and reads
\begin{eqnarray}
\mbox{\rm tr}\left\{ X(k,n)\right\} &=&
\frac{1}{f(n)}\frac{1}{2i\pi}\oint\frac{d\zeta}{\zeta}\
\left\{ \zeta ^n\overline{m}\mu _{11}^{+}\mu
_{21}^{+}-\zeta ^{-n}m\mu _{12}^{-}\mu _{22}^{-}+(p+\overline{p})\mu
_{12}^{-}\mu _{21}^{+}\right.\nonumber\\
&&\qquad\left.-(\omega _1^{+}-\omega _2^{+})\mu _{12}^{+}\mu
_{21}^{+}+(\omega _1^{-}-\omega _2^{-})\mu _{12}^{-}\mu _{21}^{-}\right\}.
\end{eqnarray}
Due to the analyticity of the function
\[
\frac1{k}(\omega _1(k)-\omega _2(k))\mu _{12}(k)\mu _{21}(k)
\]
inside and outside of the circle the last two terms in the r.h.s. vanish
and we obtain for the evolution equation of $f(n)$
\begin{equation}
\label{butlast}
f_t(n)=\int_{-\pi}^{+\pi}d\theta\
\left\{ \zeta ^n\overline{m}\mu _{11}^{+}\mu
_{21}^{+}-\zeta ^{-n}m\mu _{12}^{-}\mu _{22}^{-}+(p+\overline{p})\mu
_{12}^{-}\mu _{21}^{+}\right\}.
\end{equation}
This result can also be expressed in terms of the physical quantities $A_j$
and $I_j$ by inverting (\ref{exp-mu-base}) to get on the unit circle
\begin{eqnarray*}
\mu _{11}^{+} &=&\overline{\mu }_{22}^{\,-}=\frac{A_1\overline{I}_1+%
\overline{A}_2I_2}{|I_1|^2+|I_1|^2} \\
\mu _{21}^{+} &=&-\overline{\mu }_{12}^{\,-}=-\zeta ^{-n}\frac{\overline{A}%
_1I_2-A_2\overline{I}_1}{|I_1|^2+|I_1|^2}
\end{eqnarray*}
and then by inserting these formulae and those for
$p+\overline{p}$ and $m$ in (\ref{p-de-k}) and (\ref {m-de-k}) into
(\ref{butlast}). We obtain finally
\begin{equation}
f_t(n)=\frac 1{2\pi }\int_{-\pi }^\pi d\vartheta \,(g+\overline{g})\frac{%
|I_1|^2|A_2|^2-|I_2|^2|A_1|^2}{|I_1|^2+|I_2|^2}.
\end{equation}
Note that $f(n)$ is conserved if $g$ is pure imaginary.
\appendix

\section{Mathematical tools}

\subsection{Basic distributions}

The distributions $\delta ^{\pm }(\lambda ,1)$ and $\delta (\lambda ,1)$
have support on the circle ${\cal C}$ of radius 1 in the complex $\lambda $%
-plane and are defined by the following formulae
\begin{eqnarray}
\int \!\!\!\!\int_{{\cal D}}d\lambda \wedge d\overline{\lambda }\,\delta
^{\pm }(\lambda ,1)f(\lambda ) &=&\oint_{{\cal C}}d\zeta \,f\left( (1\mp
0)\zeta \right)  \label{delta-c1} \\
\int \!\!\!\!\int_{{\cal D}}d\lambda \wedge d\overline{\lambda }\,\delta
(\lambda ,1)f(\lambda ) &=&\oint_{{\cal C}}d\zeta \,f(\zeta )
\end{eqnarray}
or, equivalently, by
\begin{eqnarray}
\int \!\!\!\!\int_{{\cal D}}d\lambda \wedge d\overline{\lambda }\,\delta
^{\pm }(\lambda ,1)f(\lambda ) &=&i\int_{-\pi }^\pi d\theta e^{i\theta
}f\left( (1\mp 0)e^{i\theta }\right)  \label{delta-dirac} \\
\int \!\!\!\!\int_{{\cal D}}d\lambda \wedge d\overline{\lambda }\,\delta
(\lambda ,1)f(\lambda ) &=&i\int_{-\pi }^\pi d\theta e^{i\theta }f\left(
e^{i\theta }\right) .
\end{eqnarray}
Note that the distributions $\delta ^{\pm }(\lambda ,1)$ can operate on
functions which have defined left or right limit on ${\cal C}$, while the
distribution $\delta (\lambda ,1)$ can operate on functions continuous on $%
{\cal C}$ or with support on ${\cal C}$.

By complex conjugation of the equation (\ref{delta-dirac}) and by the change
of variable $\theta \to -\theta $, we obtain
\begin{equation}
\int \!\!\!\!\int_{{\cal D}}d\lambda \wedge d\overline{\lambda }\,\overline{%
\delta ^{\pm }(\lambda ,1)}f(\lambda )=i\int_{-\pi }^\pi d\theta e^{i\theta
}f\left( (1\mp 0)e^{-i\theta }\right) .  \label{delta-conj}
\end{equation}
Next, by means of the change of variable $\lambda \to 1/\lambda $ (remember
that the domain ${\cal D}$ does not contain the point $\lambda =0$) and $%
\theta \to -\theta $, we obtain
\begin{equation}
\int \!\!\!\!\int_{{\cal D}}d\lambda \wedge d\overline{\lambda }\,\delta
^{\pm }({\frac 1\lambda },1)f(\lambda )=i\int_{-\pi }^\pi d\theta e^{i\theta
}f\left( (1\pm 0)e^{-i\theta }\right) ,  \label{delta-inv}
\end{equation}
and consequently
\begin{equation}
\delta ^{\pm }({\frac 1\lambda },1)=\overline{\delta ^{\mp }(\lambda ,1)}.
\label{sym-prop-1}
\end{equation}
Now, through the change of variable $\lambda \to \overline{\lambda }$, we
obtain
\begin{equation}
\int \!\!\!\!\int_{{\cal D}}d\lambda \wedge d\overline{\lambda }\,\delta
^{\pm }(\overline{\lambda },1)f(\lambda )=i\int_{-\pi }^\pi d\theta
e^{i\theta }f\left( (1\mp 0)e^{-i\theta }\right) .  \label{delta-bar-arg}
\end{equation}
which implies the second symmetry property
\begin{equation}
\delta ^{\pm }(\overline{\lambda },1)=\overline{\delta ^{\pm }(\lambda ,1)}.
\label{sym-prop-2}
\end{equation}
These two relation naturally leads to
\begin{equation}
\delta ^{+}(\lambda ,1)=\delta ^{-}({\frac 1{\overline{\lambda }}},1).
\label{sym-prop-3}
\end{equation}
Similar symmetry properties can be obtained for $\delta (\lambda ,1)$.

\subsection{The generalized $\overline{\partial }$-formula}

Let $F^{+}\in C^1(\overline{D^{+}})$ where $D^{+}$ is the open disk of
radius 1 centered in the origin and $F^{-}\in C^1({\cal C}D^{+})$, let $%
F^{+} $ and $F^{-}$ satisfy the H\"{o}lder condition on the circle ${\cal C}$
of radius 1 and let $F^{-}$ vanish at large $z$. Then by noting $F(z)$ the
function defined as $F^{+}(z)$ for $z\in D^{+}$ and as $F^{-}(z)$ for $z\in
D^{-}\equiv {\cal C}\overline{D^{+}}$ we have
\begin{eqnarray}
&&F(z)=\frac 1{2\pi i}\oint_{{\cal C}}\frac{F^{+}(\zeta )-F^{-}(\zeta )}{%
\zeta -z}d\zeta +\frac 1{2\pi i}\int \!\!\!\int_{D^{+}}\frac{\partial
F/\partial \overline{\lambda }}{\lambda -z}d{\lambda }\wedge d\overline{%
\lambda }+  \nonumber \\
&&\frac 1{2\pi i}\int \!\!\!\int_{D^{-}}\frac{\partial F/\partial \overline{%
\lambda }}{\lambda -z}d{\lambda }\wedge d\overline{\lambda }
\label{CauchyGreen}
\end{eqnarray}
where the circle ${\cal C}$ is anti clockwise oriented.

If we define the $\overline{\partial }$--derivative of a function $F(z)$
discontinuous on ${\cal C}$ as follows
\begin{equation}
\frac{\partial F}{\partial \overline{z}}=\left( F^{+}(\zeta )-F^{-}(\zeta
)\right) \delta (z,1)+\phi _{D^{+}}(z)\frac{\partial F}{\partial \overline{z}%
}+\phi _{D^{-}}(z)\frac{\partial F}{\partial \overline{z}},\quad \zeta
=\frac z{|z|}  \label{localDBAR}
\end{equation}
where $\phi _A(z)=1$ for $z\in A$ and $\phi _A(z)=0$ for $z\notin A$ the
generalized $\overline{\partial }$--formula (\ref{CauchyGreen}) can be
rewritten as
\begin{equation}
F(z)=\frac 1{2\pi i}\int \!\!\!\int_{\overline{D^{+}}\cup D^{-}}\frac{%
\partial F/\partial \overline{\lambda }}{\lambda -z}d\lambda \wedge d%
\overline{\lambda }.
\end{equation}
Formula (\ref{localDBAR}) can be considered as the local formulation of the
generalized Cauchy--Green formula (\ref{CauchyGreen}).

Subtracting formula (\ref{CauchyGreen}) at $z=a$ we obtain
\begin{eqnarray}
&&F(z)=F(a)+\frac 1{2\pi i}\int_{{\cal C}}\frac{F^{+}(\zeta )-F^{-}(\zeta )}{%
\zeta -z}\left( \frac{z{-}a}{\zeta {-}a}\right) d\zeta +  \nonumber \\
&&\frac 1{2\pi i}\int \!\!\!\int_{D_{a,0}^{+}}\frac{\partial F/\partial
\overline{\lambda }}{\lambda -z}\left( \frac{z{-}a}{\lambda {-}a}\right) d{%
\lambda }\wedge d\overline{\lambda }+\frac 1{2\pi i}\int
\!\!\!\int_{D^{-}}\!\!\frac{\partial F/\partial \overline{\lambda }}{\lambda
-z}\left( \frac{z{-}a}{\lambda {-}a}\right) d{\lambda }\wedge d\overline{%
\lambda }  \label{asubtracted}
\end{eqnarray}
if $a\in D^{+}$ and an analogous formula if $a\in D^{-}$. The second
integral on the right and side is obtained first by computing it on the set $%
D_{a,\epsilon }=\{\lambda :\lambda \in D,|\lambda -a|>\epsilon \}$ and then
by taking the limit $\epsilon \to 0$. Note that the formula remains valid
also if $F(z)$ is going to a constant different from $0$ for $z\to \infty $.

If for $z\to \infty $ $F(z)\to F(\infty )$ we can apply (\ref{CauchyGreen})
to $F(z)-F(\infty )$ getting
\begin{eqnarray}
&&F(z)=F(\infty )+\frac 1{2\pi i}\int_{{\cal C}}\frac{F^{+}(\zeta
)-F^{-}(\zeta )}{\zeta -z}d\zeta +  \nonumber \\
&&\frac 1{2\pi i}\int \!\!\!\int_{D^{+}}\frac{\partial F/\partial \overline{%
\lambda }}{\lambda -z}d{\lambda }\wedge d\overline{\lambda }+\frac 1{2\pi
i}\int \!\!\!\int_{D^{-}}\frac{\partial F/\partial \overline{\lambda }}{%
\lambda -z}d{\ \lambda }\wedge d\overline{\lambda }.  \label{inftysubtracted}
\end{eqnarray}

Finally let us note that the Sokhotski-Plemelj formula on the circle reads
\begin{equation}
\oint_{{\cal C}}{\frac{d\zeta ^{\prime }\,}{\zeta ^{\prime }-(1\mp 0)\zeta }}%
\,f(\zeta ^{\prime })=\pm i\pi f(\zeta )+P\!\!\oint_{{\cal C}}{\frac{d\zeta
^{\prime }}{\zeta ^{\prime }-\zeta }}\,\,f(\zeta ^{\prime }),\quad |\zeta
|=1,  \label{sok-plem}
\end{equation}
where $P\!\!\oint $ denote the Cauchy principal value integral.

\subsection{Limits of integrals}
Let us proof that
\begin{equation}
\lim_{n\rightarrow \infty }\frac 1{2\pi i}P\oint \frac{d\zeta }{\zeta -k}%
\left( \frac k\zeta \right) ^n\Phi (\zeta )=-\frac 12\Phi (k),\qquad |k|=1.
\end{equation}
Under the following successive changes of variables
\begin{equation}
\zeta =e^{i\vartheta },\quad k=e^{i\varphi },\quad
\alpha =\vartheta -\varphi,\quad
x=n\alpha
\end{equation}
we derive
\begin{equation}
\frac 1{2\pi i}P\oint \frac{d\zeta }{\zeta -k}\left( \frac k\zeta \right)
^n\Phi (\zeta )=\frac 1{4\pi i}P\int_{-(\pi +\varphi )n}^{(\pi -\varphi
)n}dx\frac{e^{-ix}}{n\sin (x/2n)}e^{ix/2n}\Phi (e^{i(x/n+\varphi )})
\end{equation}
and taking the limit, for $-\pi <\varphi <\pi $,
\begin{eqnarray*}
\lim_{n\rightarrow \infty }\frac 1{2\pi i}P\oint \frac{d\zeta }{\zeta -k}%
\left( \frac k\zeta \right) ^n\Phi (\zeta ) &=&\frac 1{2\pi i}P\int_{-\infty
}^\infty dx\frac{e^{-ix}}x\Phi (e^{i\varphi }) \\
&=&-\frac 1{2\pi }\int_{-\infty }^\infty dx\frac{\sin x}x\Phi (e^{i\varphi
})=-\frac 12\Phi (e^{i\varphi }).
\end{eqnarray*}

\end{document}